\documentclass[usenatbib,twocolumn]{mn2e}
\usepackage{graphicx}
\usepackage{amsmath}
\usepackage{amssymb}
\usepackage{bm}
\usepackage{physics}
\usepackage{subfigure}
\newcommand{\beq}{\begin{equation}}
\newcommand{\eeq}{\end{equation}}

\def\bar{\overline}

\def\OB{\overline{\bf B}}

\def\emfb{\overline{\mbox{\boldmath ${\cal E}$}} {}}

\def\beq{\begin{equation}}
\def\ee{\end{equation}}
\def\lsim{\mathrel{\rlap{\lower4pt\hbox{\hskip1pt$\sim$}}
    \raise1pt\hbox{$<$}}}
\def\gsim{\mathrel{\rlap{\lower4pt\hbox{\hskip1pt$\sim$}}
    \raise1pt\hbox{$>$}}}

\def\bbJ{\bar {\bf J}}

\def\OB{\overline{\bf B}}

\def\tted{\tilde{\tau}_{ed}}
\def\ttr{\tilde{\tau}_r}
\def\tro{\tilde{Ro}}
\def\ty{\tilde{y}}
\def\tilf{\tilde{f}}
\def\tr{\tilde r}

\title[Some consequences of shear on   dynamos]
 {Some  consequences of  shear   on galactic dynamos with   helicity fluxes}
\author [Zhou \& Blackman]
{Hongzhe Zhou $^{1}$\thanks{E-mail: hzhou21@ur.rochester.edu}, Eric G. Blackman$^{1,2,3}$\thanks{E-mail: blackman@pas.rochester.edu}]\\
 $^{1}$Department of Physics and Astronomy, University of Rochester, Rochester NY, 14627, USA\\
 $^{2}$Laboratory for Laser Energetics,  University of Rochester, Rochester NY, 14623, USA\\
 $^{3}$Kavli Institute for Theoretical Physics,  UC Santa Barbara, Santa Barbara, CA, 93106, USA\\}

\begin{document}

\date{}
\pagerange{\pageref{firstpage}--\pageref{lastpage}} \pubyear{}
\maketitle
\label{firstpage}
\begin{abstract}
Galactic dynamo models sustained by supernova (SN) driven turbulence and differential rotation have   revealed  that  the sustenance of  large scale fields requires a flux  of small scale magnetic helicity to be viable.
 Here we  generalize a  minimalist analytic version of such galactic dynamos to  explore some heretofore unincluded contributions from  shear on the total turbulent energy and  turbulent correlation time, with the helicity fluxes maintained by either winds, diffusion,
 %EB4 added diffusion
  or magnetic buoyancy.
  We construct an analytic framework for modeling the turbulent energy and correlation time  as function of SN rate and  shear.  We compare our prescription  with previous approaches that only include rotation.  The solutions depend
  separately on the rotation period and the eddy turnover time and not just on their ratio (the Rossby number).
 We consider models in which these two time scales are allowed to be independent and also a case in which they
 are mutually dependent on  radius when a radial  dependent SN rate model is invoked.
For the  case of a fixed rotation period (or fixed radius)  we show that the influence of shear is  dramatic  for low Rossby numbers,   reducing
the correlation time of the turbulence, which in turn, strongly reduces the saturation value of the dynamo compared to the case when the shear is ignored.
We also   show that
even in the absence of winds or diffusive fluxes, magnetic buoyancy may be able to sustain sufficient helicity
fluxes to avoid quenching.

\end{abstract}

\section{Introduction}

\subsection{Background}
In situ galactic dynamo theory has long been a leading paradigm to explain the ordered large scale magnetic fields of galaxies \citep{rss1988}.
In this paradigm, a  weak seed field, perhaps supplied primordially, is   amplified via the
action of turbulence and differential rotation in the galactic interstellar medium.
How such dynamos work in detail, has been a longstanding research enterprise \citep{rss1988,brandenburg2005,Shukurov2006,Hanasz2009,chamandy2014,Blackman2015SS,Kulsrud2015}.

Standard (20th century)  mean field $\alpha-\Omega$ galactic dynamos
typically  have at least three key ingredients (1) supernova driven turbulence, which in the presence of galactic rotation and stratification produces a kinetic helicity driven  "$\alpha$" effect
that converts toroidal to poloidal field
and (2) differential rotation that shears the poloidal
field into the toroidal direction and (3) some kind of  turbulent
diffusion or loss term of the mean field in a thin disk that ensures the
that the net toroidal flux in the disk
reflects the observed field geometry (e.g. quadrupole).

A  challenge for   20th century galactic dynamo theory
has been  the absence of a physical understanding of how the dynamo saturates. That basic  theory is kinematic, considering only the growth of the large scale field without including the dynamics of the field on the driving flow.  Intertwined with this deficiency has been
the realization that  standard mean field textbook $\alpha-\Omega$ dynamos also do not conserve magnetic helicity \citep{Blackman2000,Vishniac2001}. (For  reviews
see \cite{brandenburg2005} and \cite{Blackman2015SS}).

Principles of dynamically including magnetic helicity
conservation in MHD turbulence from  \cite{Pouquet1976} and modified lessons from steady-state mean field considerations of \cite{Gruzinov1994} and \cite{Bhattacharjee1995}
were synthesized into time-dependent mean field dynamical toy models \citep{Blackman2002} using a simple closure (now referred to as "minimal $\tau$").  In these models, the growth of a helical component of the large scale field is accompanied by growth of the oppositely
signed small scale helical field which in turn, represents a backreaction that saturates the dynamo.
For dynamos without shear this leads to a steady state, but for dynamos with shear,
this can lead to catastrophic quenching alleviated only when helicity fluxes carry away the excess small scale field. Ultimately this requires  a dynamo sustained by helicity fluxes (\cite{Blackman2000}).
Depending on which terms in the electromotive force actually dominate, a  complementary
perspective is that the large scale dynamo is sustained directly via helicity fluxes even in the absence of any kinetic helicity (e.g. \cite{Vishniac2001,Vishniac2014}).
Helicity flux driven dynamos are conceptually related to
the sustenance of large scale fields in the different context of laboratory
magnetically dominated plasmas \citep{Strauss1985,Bhattacharjee1986}.

Incorporation  of some these  principles
has led the numerical demonstration of the helpful role of magnetic helicity in numerical simulations
of dynamos in stellar contexts \citep{Brandenburg2004,Chatterjee2011}
as well as  practical   galactic dynamo models with helicity fluxes
 \citep{Shukurov2006,sur2007,Chamandy2016}.

% In galaxies, due to the large magnetic Reynolds numbers, magnetic fields lines are frozen into the %streamlines.
%EB ironically, the large RM also causes turbulence which in turn leads to  high magnetic diffusivity and low effective RM...
%stretches poloidal field lines in the azimuthal direction,  while turbulence in the interstellar medium (ISM) will entangle field lines and tend to line them up in the poloidal direction

A second challenge of  galactic and mean field dynamo theory is to incorporate the  influences  of rotation and shear on the  turbulence, dynamo coefficients, and EMF.
One  approach is to expand the
turbulent quantities into a base state that is independent of shear and rotation
plus corrections that depend on them. The resultant
mean turbulent EMF (whose curl enters the growth if the mean magnetic field)  can then be expanded  into a sum of all possible terms that are linear in the mean magnetic field and linear in the mean rotation or shear \citep{brandenburg2005,Radler2006}.  The relevance and interpretation of each of these terms must be assessed independently for a given circumstance.
However, this approach does not capture all of the effects of rotation and shear
to all orders. Doing so formally is  impractical,  but physical approaches can provide
insight and shortcuts.
%In this paper we take such an approach to explore influences of shear and rotation on the correlation time  and  energy input rate of the turbulence.

\subsection{Strategy and Outline}
The influence of rotation can be partly gauged by the ratio of the nonlinear term in the Navier
Stokes equation to the Coriolis term in the rotating frame.
This dimensionless ratio, the Rossby number,  is given by
%HZ6 ref:
\beq
Ro(\tau_{ed})= {\tilde Ro}({\Omega})=\frac{1}{\Omega\tau_{ed}},
\label{1Ro}
\eeq
where  $\Omega$  is the rotation speed and $\tau_{ed}$ is the eddy turnover time,
presently defined in terms of the turbulence supplied specifically by supernovae.
The latter is important to keep in mind as we will also utilize a separate correlation time $\tau_{cor}$ not  necessarily equal to $\tau_{ed}$.
The above equation introduces our convention of writing
$Ro$ for the Rossby number for fixed $\Omega$ allowing $\tau_{ed}$ to vary
and $\tilde Ro$ for the Rossby number at fixed $\tau_{ed}$ allowing $\Omega$ to vary.
How dynamos
 depend separately on  $\tau_{ed}$, $\Omega$  and on differential rotation
is not completely  understood. Even for the basic $\alpha-\Omega$ type
dynamo, the question of how  the kinetic component of the helicity coefficient
%HZ7 the kinetic component of the helicity coefficient?
%EB7 yes, fixed..
contribution $\alpha_0$ depends on rotation and shear warrants revisiting for strong shear.
%EB define alpha here probably.
%EB refs.
%, due to the complexity of turbulent motions and a lack of dynamical theories relating the correlation time o$\tau_{cor}$ (and therefore the kinetic helicity $\alpha_0$) to $
%EB we are also assuming tau_c is same for all coefficients but not always true also in the minimal tau approach for dynano coefficients...its not quite a correlation time that appears..hmm.
%\Omega$, $\tau_{ed}$, and other parameters.
%Specifically,  standard calculations of the dynamo transport coefficients depend on the
%correlation times of quadratic functions of turbulent quantities, typically assumed to be a
%constant $\tau_c$ for all coefficients.

There are a few precursors in this context.
\cite{ruediger1978}  calculated an effect of rotational quenching on  $\alpha$.   \cite{rss1988}   considered the effect of the Coriolis force without shear and their prescription
for the effect of rotation on $\alpha_0$ can be recast by replacing the correlation time of the turbulence $\tau_c=\tau_{ed}Ro^{-1/2}$ when $Ro\geq 1$, and
$\tau_c=\tau_{ed}$ otherwise. In \cite{chamandy2016statistical}, the same resulting piecewise-defined $\alpha$ was used.
\cite{blackman2015} and \cite{Blackman2016} included an  effect of shear on the correlation time by
arguing that
 $\tau_{cor}$ equals  $\tau_{ed}$ times  a factor that depends on $Ro$ and shear.

In the present paper we   explore and  generalize a physical model  for the influence
of shear and rotation on both  $\tau_{cor}$ and the turbulent energy density   for galactic dynamos.
%combining the ideas of \cite{rss1988} and \cite{blackman2015} and making generalizations upon that.
%Revisiting some of the arguments made in the stellar context  \cite{blackman2015} is a helpful
%starting point in building our model.
 %EB having said the above
 %Shear contributes both to
%oth the shear (characterized by $\Omega$) and the eddy (characterized by $\tau_{ed}$) in the kinematic regime, following by a physical argument.
%both  high Rossby number indicates a galaxy having relatively strong turbulence or slow rotation, which means different parts within an eddy will be notably correlated until a second eddy occurs nearby, mixing everything and rebuilding the correlation. Therefore the lifetime of an eddy is mostly determined by $\tau_{ed}$ in this case. On the other hand, for $R_o\ll1$ which implies weak turbulence or strong rotation, eddies will have been shredded by shear before new ones are established, so
%$\tau_{cor}\approx\tau_s$ with $\tau_s$ being the characteristic time of the differential rotation and to be clarified in the next section.
We will see that when  $Ro,{\tilde Ro}>>1$, the supernova turbulence dominates both the turbulent energy density and its correlation time.  In the regime $Ro,{\tilde Ro}<<1$ shear can dominate both of these quantities.  We build our model in three  separate ways, first  fixing $\Omega$ and changing $\tau_{ed}$, and then  fixing $\tau_{ed}$ and changing $\Omega$. Then we  consider a model
in which they are mutually dependent on radius, via  their connection to the star formation rate.
%HZ 1/5
%To distinguish the two cases, we denote the Rossby number by $Ro$ in the former case and by $\tro$ in the latter case.
%
The need for this arises because the dynamo depends separately on those two parameters not just in their dimensionless combination of the Rossby number. We explicitly derive  $\tau_{ed}$ in terms of the SN rate and show how $\tau_{cor}$
changes as a function of rotation and shear.
Both the effect of shear on the turbulent correlation time and as a supplemental source
of turbulent energy have not been included in galactic dynamo models, although
in  the absence of SN, shear is expected to be a source of galactic   turbulence \citep{BalbusSellwood1999}.
%because in galaxies about ninety percent of the energy of ISM turbulent motions comes from supernovae explosions \citep{rss1988}.
%The characteristic time scale of these events, or eddy turnover time $\tau_{ed}$, are shorter in young galaxies than that in old ones.
%Another reason for having various values of $\tau_{ed}$ could be having different mechanisms of turbulence. For instance, turbulent energy can be extracted from differential rotation when magnetic fields of suitable strength exist, as suggested by .

We also incorporate a magnetic buoyancy (MB) term  \citep{parker1966} in the helicity flux term of the dynamo  equations, generalizing the corresponding therm of \cite{sur2007} which included only an advective wind flux term.
% In most dynamo models only the stellar wind is taken into account, appearing as an independent parameter of the magnetic fields. Here we consider in addition a buoyancy term caused by Parker instability if horizontal magnetic field lines inside the ISM are perturbed vertically, gas tends to 'slide' down along the magnetic field lines, causing a lower matter density in the region where it was, and making it buoyant.
The buoyant speed  itself depends on the magnetic field, which increases the nonlinearity of the dynamo equations.

The paper is arranged as follows. In Sec. \ref{sec2} we relate the turbulent velocity and  correlation time to the Rossby number in both fixed-$\tau_{ed}$ and fixed-$\Omega$ cases, developing expressions for both the turbulent energy density and correlation time as a function of shear, rotation, and SN rate.
For a given shear profile we consider three cases: (i) fixed $\Omega$, varying $\tau_{ed}$; (ii) fixed $\tau_{ed}$, varying $\Omega$;  and (iii) mutually dependent variation of $\Omega$ and $\tau_{ed}$.
We apply these relations to the dynamo equations in Sec. \ref{sec3}. The solutions are found numerically in Sec. \ref{sec4}, where we show both steady state solutions and time evolution of the magnetic fields.
We identify where the results from our calculations that include the new ingredients
differ from  previous approaches. We also discuss the influence of magnetic buoyancy and the consequences of our calculations for observed pitch angle. We conclude in Sec. \ref{sec5}.

%2/2 review
\section{Effect of shear on correlation time and  turbulent energy }\label{sec2}
The Rossby number is  function of  two variables $(\tau_{ed},\Omega)$.
The value of  $\tau_{ed}$  can vary for different  supernova rates and $\Omega$
  depends on the details of galaxy formation and the mass therein. In practice, these two quantities could be correlated because a fixed initial mass function for stars, and a baryon mass correlated
 with total mass would increase both the rate of SN and the rotation rate at a fixed
 radius.   Below we consider separately cases where we allow these two quantities to be independent and then
 consider a case where they are mutually dependent.
When they are independent, the dynamo then depends  on these two  variables independently, not just their ratio.

We first construct a physical model for the influence of shear on the turbulent energy
and correlation time  by fixing $\Omega$ and allowing $\tau_{ed}$ to change.
We then construct the analogue where we keep  $\tau_{ed}$ and allowing $\Omega$ to change.
We show in the appendix that these two approaches can be unified.   In the last subsection of this section
we consider the case where the two quantites are mutually dependent.

%After some key ideas in the derivation become familiar, it will be straightforward to generalize the model and make both $\tau_{ed}$ and $\Omega$ free parameters. We illustrate how it works in the last part of this section.
In what follows, quantities with a subscript $0$ (e.g., $\tau_{ed0}$, $v_0$, $l_{ed0}$ and so on)
are evaluated at their fiducial values such that
 $Ro_0=\tro_0=(\tau_{ed0}\Omega_0)^{-1}=1$.

\subsection{Correlation time and turbulent energy for fixed $\Omega$, fixed shear, but different SN rates}\label{2.1}
\subsubsection{Effect of the shear on the correlation time}
\label{2.1.1}
We  distinguish between the turbulent correlation time $\tau_{cor}$  and the naked eddy turnover
time $\tau_{ed}$ determined by  SN in the  absence of shear, and $\tau_{ed0}$ as the fiducial value of the latter.  We define the ratio of the former to latter  as
 %relation between $\tau_{cor}$ and $\tau_{ed}$ by first giving several constraints on
\beq\label{tc}
y(Ro)
\equiv\tau_{cor}/\tau_{ed0},
\eeq
where $Ro=Ro(\tau_{ed})$  for fixed $\Omega$ in this section.
The quantity  $\tau_{cor}$
 must satisfy the  physically expected behaviors in the low and high $Ro$ limits,
  namely  $\tau_{cor}\rightarrow\tau_{ed}$ as $Ro\rightarrow\infty$ and $\tau_{cor}\rightarrow \tau_s$ as $Ro\rightarrow0$,  where $\tau_s$ is defined as
\beq
\tau_s=\frac{\Delta r}{r\Delta\Omega}=\frac{\Delta r}{r\Delta r\partial_r\Omega}=\frac{1}{q\Omega}=\frac{\tau_{ed0}}{q}
\eeq
along with the rotation profile $\Omega\propto r^{-q}$. The physical meaning of $\tau_s$ is evident if we consider radially separated points on  two  concentric rings orbiting in  the galaxy with radii $r-\Delta r/2$ and $r+\Delta r/2$ respectively. Their relative velocity will be $r\Delta\Omega=r\Delta r\partial_r\Omega$, and  $\tau_s$ characterizes the time scale for these points to further separate  to
by $\Delta r$ in the azimuthal direction.
In  terms of $y$, the aforementioned asymptotic limits imply
\beq\label{cst1}
y=\left\{
{\begin{aligned}&1/q\ \ &Ro\rightarrow0\\
&1/Ro\ \ &Ro\rightarrow\infty
\end{aligned}}
\right\}.
\eeq

Deriving $y$ from first principles  is a  challenging endeavor but
we can make good progress with a physically motivated approach.
We posit that quadratic time correlations of turbulent quantities decay exponentially in time
over a correlation time
that  has separate independent exponential factors from
shear and SN turbulence. Then
\beq\label{yexp}
\tau_{cor}^{-1}=\tau_{ed}^{-1}+\tau_s^{-1},
\eeq
or  equivalently,
\beq\label{y}
y=\frac{1}{Ro+q}.
\eeq
Eq. (\ref{y}) satisfies the constraint (\ref{cst1}).
In the fast rotation limit $Ro\rightarrow0$, we have $y=1/q$
%EB removed \rightarrow0$
so that Eq. (\ref{cst3}) predicts a correlation time that asymptotically approaches  a constant for $q>0$, as we will see below.
%rather than monotonically increases as $Ro$ decreases.
%This comes as a result of relating $\tau_{cor}$ to shear and highlights it importance and necessity.

\subsubsection{Effect of the shear on the turbulent energy}
Next, we consider the effect of the shear on the turbulent energy. Technically the turbulent energy consists of both energy from supernovae and differential rotation
since rotating MHD shear flows with $q>0$ are unstable \citep{velikhov1959,Balbus1991}.
%HZ6
%EB references to be added.
 In terms of energy density input rate, this implies
\beq\label{turbngy}
\frac{\rho v^2}{\tau_{cor}}=\frac{E}{\tau_{ed}l_{ed}^3}+\frac{\varepsilon}{\tau_s}
\eeq
where $\rho$ is the average density of ISM, $v$ is the mean square root velocity of the turbulence, $E$ is the energy input to the ISM per supernova, $l_{ed}=v\tau_{ed}$ is the eddy scale, and $\varepsilon$ is the energy density input by shear and is taken to be a fraction
%HZ edited below on 12/14
of the fiducial shear energy density $\rho v_{s0}^2=\rho (l_{ed0}/\tau_s)^2$.
More specifically,  we then have
%EB edited below
\beq
\varepsilon=\xi \rho (l_{ed0}/\tau_s)^2=\xi \rho l_{ed0}^2 q^2 \tau_{ed0}^{-2}=\xi q^2\rho v_0^2
\label{shear}
\eeq
%HZ edited above on 12/14
where we take $\xi=0.1$.
To provide physical meaning for the second term in Eq. (\ref{turbngy}), we note that  the energy density supplied by SN per unit time can be expressed as $E\Gamma V^{-1}$ where $V$ is the volume of the galaxy, and $\Gamma$ the rate at which SNe are produced in $V$.  Crudely assuming SN occur isotropically, we have
\beq
\frac{\Gamma}{\tau_{ed}^{-1}}\simeq \frac{V}{l_{ed}^3}
\eeq
where $l_{ed}^3$ indicates the turbulent correlation scale from SN.
%EB is my change ok above
Therefore
\beq
\frac{E\Gamma}{V}=\frac{E}{\tau_{ed}l_{ed}^3}.
\eeq
We further assume that $E$ is a constant, and that the variation of $\rho^{1/3}$ with $Ro$ is small compared to $v$, so that $\rho$ can be taken as approximately  constant as well.
For the fiducial point values, the ratio between the second term on the RHS to the LHS of Eq. (\ref{turbngy}) is
\beq
\frac{0.1\rho v_0^2/\tau_s}{\rho v_0^2/\tau_{cor}}=\frac{0.1\rho v_0^2/(\tau_{ed0}/q)}{\rho v_0^2/(\tau_{ed0}y)}=\xi q^3 y=\frac{1}{20}
\eeq
in using (\ref{tc}) and (\ref{y}).
For a flat rotation profile (as in typical spiral  galaxies),
$q\simeq 1$ so that this ratio is small
and at the fiducial point values we can neglect the second term on the RHS of Eq. (\ref{turbngy}) to obtain
% energy at the fiducial point is  dominated by the SN,
%EB the above comment seems circlular since we assumed that  the shear energy was 0.1 of the turbulent energy so it seems to be
\beq
E/\rho\simeq v_0^5\tau_{ed0}^3
\eeq
which can then be used to simplify  Eq. (\ref{turbngy}) to
\beq\label{feqn}
\frac{f^2}{y}=\frac{Ro^4}{f^3}+\frac{q^3}{10},
\eeq
where
 \beq\label{v}
f(Ro)\equiv v/v_0.
\eeq

Eq. (\ref{feqn}) determines the nonlinear relation between the turbulent speed $v$ and
 $Ro$. However its solution has  simple asymptotic behaviors. In the large $Ro\gg1$ regime, $\tau_{ed}\rightarrow0$ and SN energy input rate
 %EB we have to be careful about energy vs. input rate
 dominates over that of the shear, so we can drop the second term in Eq. (\ref{feqn}) which leads to $f=(Ro^4y)^{1/5}$. In the $Ro\ll1$ regime, the second term of Eq.  (\ref{feqn}) dominates so we drop the first term the right to obtain get $f=(q^3 y/10)^{1/2}$. The two terms contribute equally  at $Ro=0.22$ so we approximate $f$ as
\beq\label{f}
f=\left\{\begin{aligned}&(Ro^4y)^{1/5}\ \ &Ro\geq0.22\\&(q^3 y/10)^{1/2}\ \ &Ro\leq0.22\end{aligned}
\right\}.
\eeq
These relations capture the fact that  as  SN become scarce, the average turbulent speed of the ISM would decrease, but since shear provides  a fixed baseline of turbulent energy, $v$ approaches to a constant.

Given Eq. (\ref{f}) we are poised to check one more  plausibility condition for  $y$, namely that the magnitude of $\alpha_0\sim\tau_{cor}\langle \vec{v}\cdot\nabla\times\vec{v}\rangle$ cannot be larger than $v$, since the helical fraction cannot exceed unity.
%Mathematically, the root of the inequality roots in the fact that $\vec{v}\cdot\nabla\times\vec{v}\leq v^2/l_{cor}$ where $l_{cor}=v\tau_{cor}$.
For quasi-isotropic turbulence  \citep{durney1982,rss1988}
%HZ6
%EB we should add also another reference for galactic estimate..
\beq
\alpha_0\sim\xi_\alpha \tau_{cor}^2v^2\Omega/h
\eeq
where $\xi_\alpha$ is a factor smaller than one and $h\propto v^{-1/2}$ is half of the scale height of the galaxy in an isothermal self-gravitating slab model \citep{Spitzer1981}.
%HZ6
%EB need ref here
The required inequality is then   $\tau_{cor}^2v^2\Omega/h\leq v$,
%We will come back to check this inequality later after we derive the dependence of $v$ on $Ro$A second $y$ using $\alpha_0\leq v$.
or equivalently,
\beq\label{alessthanv}
y^2 f^{3/2}\leq\frac{h_0}{v_0\tau_{ed0}}=5
\eeq
upon using fiducial values $h_0=0.5\ \text{kpc}$, $\tau_{ed0}=10^{15}\ \text{s}$, and $v_0=10^6\ \text{cm/s}$, the validity of which can  checked  using  Eqs. (\ref{y}) and (\ref{f}).
%EB wanted to make sure that this is expressed not as an imposed condition but one that is naturally met by our choice.

Note also that before turbulent energy is taken over by shear as $Ro$ decreasing,
%HZ 12/14
$l_{cor}=v\tau_{cor}$ will never exceed the scale height $2h$, since
\beq
\frac{2h}{l_{cor}}=\frac{2h_0f^{-1/2}}{l_{ed0}fy}=10f^{-3/2}y^{-1}
\eeq
and it can be  verified the quantity above is always greater than unity using (\ref{f}).
%HZ 12/14 above
%EB should explain how to arrive at this previous statement with equations here
%This ensures  a self-consistent formulation.

\subsection{Correlation time and turbulent energy for fixed  SN rates but
different rotation rates}\label{2.2}
Complementing the previous subsection, here we instead
 fix the SN rate (and  thus  $\tau_{ed}$) but allow for different rotation
 rates.
 By direct analogy to (\ref{y}), we  define $\ty$ by
\beq\label{tc'}
\ty=\tau_{cor}/\tau_{ed}.
\eeq
Note that we have $\tau_{ed}=\tau_{ed0}$ here. The asymptotic limits are  now
\beq\label{cst'1}
\ty=\left\{\begin{aligned}&\tro/q\ \ &\tro\rightarrow0\\&1\ \ &\tro\rightarrow\infty\end{aligned}
\right\}.
\eeq
By analogy to Eq. (\ref{cst1}) we  take
\beq\label{y'}
\ty=\frac{\tro}{\tro+q}.
\eeq

For the turbulent energy, we  now generalize the energy input from shear to allow varying
 angular velocity, assuming a fixed fraction  is available. Thus Eq. (\ref{shear}) is replaced by (with $\xi=0.1$)
\beq
\varepsilon
%\propto\rho(\Omega r)^2 \ \Rightarrow\ \varepsilon
=\xi q^2 \rho v_0^2 \tro^{-2}
\eeq
which gives the correct value of $\varepsilon$ at the fiducial point where $\tro=Ro_0=1$. Now
%HZ 12/14
$\tilf=v/v_0$ is given by
\beq
\frac{\tilf^2}{\ty}=\frac{1}{\tilf^3}+\frac{q^3}{10\tro^3}
\eeq
of which the solution is approximately
\beq\label{f'}
\tilf=\left\{\begin{aligned}&(\ty)^{1/5}\ \ &\tro\geq0.355\\&(q^3\ty/10\tro^3)^{1/2}\ \ &\tro\leq0.355\end{aligned}\right\}
\eeq
The
plausibility analogue to Eq. (\ref{alessthanv})  becomes
\beq\label{cst'2}
\ty^2 \tilf^{3/2}/\tro\leq5,
\eeq
which is also satisfied if we use Eqs. (\ref{y'}) and (\ref{f'}), and same fiducial values $h_0=0.5\ \text{kpc}$, $\tau_{ed0}=10^{15}\ \text{s}$, and $v_0=10^6\ \text{cm/s}$ as  in the last subsection.
%HZ 12/14 above

%EB should f be \tilf in the above equation---also add a sentence to explain why it is satisfied...

%from the requirement that  $|\alpha_0|\leq v$.   $|\alpha_0|\leq v$

\subsection{Generalizing the correlation time to
include the case  of  rotation without shear}

For rigid rotation, $q\rightarrow0$, and $y=Ro^{-1}$ yields $\tau_{cor}=\tau_{ed}$
as expected in the absence of shear.
The effect of rigid rotation without shear on $\alpha_0$ has
been previously considered due to the Coriolis acceleration (see p.163 in \cite{rss1988}).
We can interpret this effect as a change  to the correlation time as follows:
%  call it effective $\tau_{cor}$ method below).
%If the internal rotational velocity incurred by an eddy from the
%Coriolis force $v_c$ exceeds that of the SN induced turbulent
%velocity, then such a parcel will shred by interaction with a neighbor faster than in the absence
%of the Coriolis force.
%EB im not sure of the above statement actually since coriolis does no work..
%would case a parcel to travel  a distance longer than the scale of an eddy,
%$l_{ed}$, in a time scale of $\tau_{cor}$,
Over a time $\Delta t$, the displacement from the Coriolis force can be estimated to be
$d \sim  \Omega v(\Delta t)^2$ and then we can  set   $\Delta t=T_c$
as the time inteval for the Coriolis force to
rotate an eddy of radius $l_{ed}/2$ by $\pi/2$, or cause a  displacement
$d=\pi l_{ed}/4$. This $T_c$ is  the time scale for two adjacent eddies to mutually shred
from only this interaction and if this is the shortest of the eddy destruction mechanisms
it would determine the correlation time.
%and so if this time scale is less than $\tau_{ed}$
%(the correlation time in the absence of rotation)
%$T_c$ becomes the actually correlation time.
 Using the above expressions for  $d$ and $\Delta t$, we obtain
%time scale  $l_{ed}/v_c $ then becomes the effective correlation time.
%Here $v_c$ is the characteristic velocity of the fluid element driven by the Coriolis force,
%The internal eddy rotation speed $v_c$  can be defined a from the acceleration incurred by the eddy center of mass  from the Coriolis force such that the effective radius of curvature is the eddy scale.
%That is,
%\beq
%v_c^2 = \Omega v l_{ed}= Ro^{-1/2}v
%\eeq
%EB re-thinking if this argument works given that Corioiis does no work, we can discuss
%and therefore the effective correlation time from only the Coriolis effect would  be
%\beq
%l_{ed}/v_c=\tau_{ed}Ro^{1/2}=\tau_{ed0}Ro^{-1/2},
%\eeq
%where we have used $\tau_{ed}/\tau_{ed0}=Ro_0/Ro$ and $Ro_0=1$.
%That $l_{ed}/v_c$ indeed can  also be interpreted as the lifetime of an eddy follows if we imagine two
%nearby eddies. The time scale for them to be shredded is the time scale for each to rotate
%an angle $\pi/2$ from $v_c$ or an azimuthal scale $l_{ed}\pi/2 \sim l_{ed}$.
% in row. The time scale for  the middle eddy to be shredded by the other two is the time of a fluid element carried by the left (or right) eddy to move across the middle one, i.e., $l_{ed}/v_c$.
%EB edited above
\beq\label{cst3a}
T_c = \left({\pi l_{ed}/4 \over \Omega v}\right)^{1/2}\simeq
\tau_{ed}Ro^{1/2}
\eeq
Combining this  with  case of Sec. \ref{2.2} (fixed $\tau_{ed}$), we can then write
%$T_c=\tau_{ed}Ro^{1/2}$ and
\beq\label{cst'3}
\frac{\tau_{cor}}{\tau_{ed}}=\text{min}\{\tro^{1/2},\ty(Ro)\}.
\eeq
For the case of Sec. \ref{2.1} (fixed $\Omega$),
but with $q=0$, we can write  $T_c=\tau_{ed0}Ro^{-1/2}$
 and then
\beq\label{cst3}
\frac{\tau_{cor}}{\tau_{ed0}}=\text{min}\{Ro^{-1/2},y(Ro)\}.
\eeq
%For rigid rotation, unlike the triviality of $y$, (\ref{cst3}) still holds . We could have chosen a different $y$ which is always smaller than $Ro^{-1/2}$ to simplify the expression, but there is no \emph{a prior} reason to do this.

\begin{figure}
{\includegraphics[width=0.9\columnwidth]{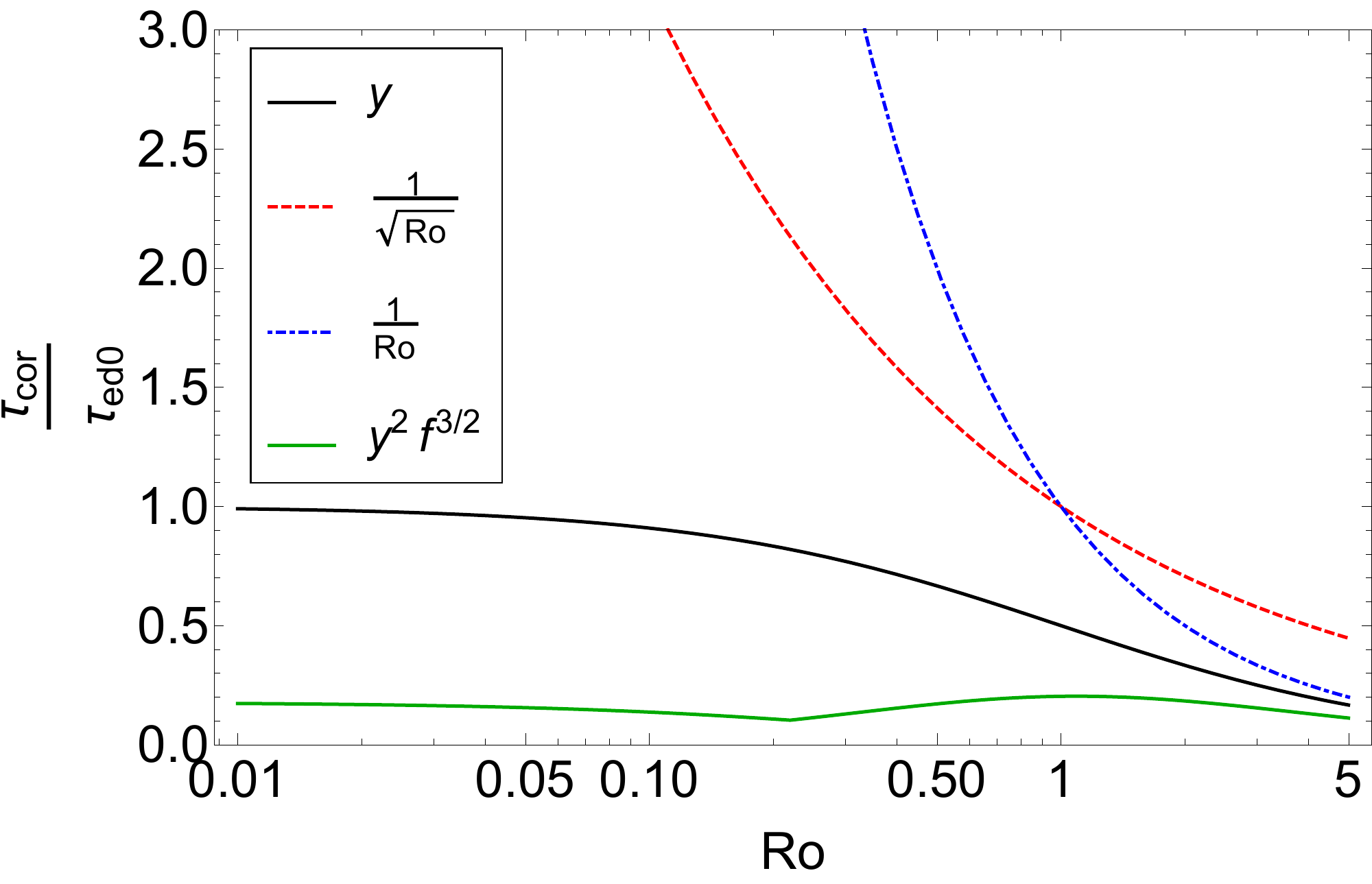}}
 \caption{Plot of $\tau_{cor}/\tau_{ed0}$ versus $Ro$ for different models, and a verification of the inequality (\ref{alessthanv}). The blue curve is for $\tau_{cor}=\tau_{ed}$, whereas the combination
 composed of the red curve below $Ro=1$ and the blue curve above $Ro=1$
  gives  the overall curve that includes only rotational quenching, $\tau_{cor}=\text{min}\{Ro^{-1/2}, Ro^{-1}\}$, without shear. Our model that includes shear defined by (\ref{y}) is given by the black curve. While it approaches $Ro^{-1}$ asymptotically as $Ro\rightarrow\infty$, there is a notable difference when $Ro$ is small, due to the effect of shear which prevents $\tau_{cor}$ from becoming arbitrarily  large without an upper bound. The green line illustrates that Eq. (\ref{alessthanv})
 is satisfied.}
  %does not  exceed 5, as required by $\alpha_0\leq v$}
 %EB maybe the two lines
 \label{tcted0}
\end{figure}

Note that the Eqs. (\ref{cst'3}) and (\ref{cst3}) incorporate the
 separate influences on the correlation time from pure rotation and shear.
Fig. \ref{tcted0} shows the correlation time  in our approach.

We may express $\tro$ as a function of the radial coordinate $r$
given the rotation profile, i.e.,
\beq
\tro=\frac{1}{\tau_{ed0}\Omega}=\frac{1}{\tau_{ed0}\Omega_0(r/r_0)^{-q}}=\left(\frac{r}{r_0}\right)^q
\eeq
where we have used $\tau_{ed0}\Omega_0=1$. Replacing $\tro$ by $r$ using the relation above
and assuming all other variables are independent of $r$
provides us with one of the simplest way to write down a $r$-dependent model.

\subsection{Case when  correlation time   and $\Omega$ {\it both} depend on $r$}
\label{secr}
%HZ Ro is not directly used in the model. What about "Important "
In the cases considered above, we have assumed that the eddy time and the rotation periods are independent
but in practice, models of star formation rates (SFR) in galaxies both depend on radius.
We now suppose that  $\tau_{ed}$ varies with the radial coordinate $r$ according the the prescription adopted by \cite{prasad2016}.
Specifically, we adopt the relation
\beq
\tau_{ed}=\frac{r}{r_0}\tau_{ed0}\propto r
\label{sken}
\eeq
where $r_0=8\ \text{kpc}$
and it is  determined from the following argument:
%represents the position of the sun in the .
 if $\tau_{ed}^{-1}$ is proportional to the SN rate and the SN rate is proportional to the surface density of the SFR $\Sigma_{SFR}$ \citep{shukurov2004,rodrigues2015}, we have  $\tau_{ed}^{-1}\propto \Sigma_{SFR}$. Further,
we assume a Schmidt-Kennicutt-like power-law relation $\Sigma_{SFR}\propto \Sigma^{\xi_g}_g$ \citep{schmidt1959,kennicutt1989,Heiderman2010},
where $\Sigma_g$ is the gas surface density and typically  $1\le {\xi_g}\le1.4$. For simplicity, we take ${\xi_g}=1$ here.
The mean galactic gas surface density $\Sigma_g\propto 1/r$ if the gas surface density hovers around a fixed fraction of order unity
  near   the critical Toomre density for  gravitational stability \citep{toomre1964,cowie1981}.
  Then combining these above relations  we arrive at Eq. (\ref{sken})
  above.
If the helicity flux is driven by a galactic fountain, which in turn is driven by SN
\citep{tenoriotagle1988,shapiro1976,shukurov2004,rodrigues2015}
,  we might  consider that the outflow speed also satisfies
%EB lets chat about this point
\beq
U\propto 1/\tau_{ed}\propto 1/r,\ \text{or},\ U=\frac{r_0}{r}U_0.
\eeq
In addition, for a flat rotation curve, $\Omega\propto 1/r$.

Now since both $\tau_{ed}$ and $\Omega$ vary with $r$, we need the unified relations derived in Appx. \ref{appx1}, which results in
\beq\label{ry}
y(r)=r/2
\eeq
and
\beq\label{rf}
F(r)=\text{max}\{(1/2r^3)^{1/5},(1/20r^2)^{1/2}\}.
\eeq

\section{Dynamo Model for Low and High Rossby Numbers}\label{sec3}
The  induction equation for the mean field is given by
(for reviews \cite{brandenburg2005,Blackman2015SS})
%HZ7
%EB7 fixed
%EB need reference
\beq
\partial_t\OB=\bm{\nabla}\times(\bar{\bm{U}}\times\OB+\emfb-\beta\bbJ)
\eeq
where $\OB$ and $\bar{\bm{U}}$ are the (ensemble or spatial averaged)
%EB  could also be spatial average
mean magnetic field and velocity field, respectively; $\bbJ=\bm{\nabla}\times\OB/\mu_0$ is the mean current (taking $\mu_0=1$); $\beta$ is the Ohmic resistive diffusion coefficient; $\emfb=\alpha\OB-\beta_t\bbJ$ is the mean turbulent electromotive force where $\beta_t$ is the turbulent magnetic diffusivity,
%EB units...  J = curl B misses the c/4pi even if mu_0=1
%HZ 12/14 fixed
 $\alpha\equiv \alpha_0+\alpha_m$ is the
pseudoscalar helicity coefficient separated into  kinetic and magnetic contributions,  $\alpha_0$ and  $\alpha_m$,  respectively.

%HZ6
%EB need refs... ill help

We adopt  cylindrical coordinates and apply the 'no-z' approximation \citep{subramanian1993,moss1995,phillips2001,sur2007} to reduce the PDEs to a simpler set of ODEs; the reduced dynamo equations read\footnote{Here we are working in the $\alpha\Omega$ dynamo approximation. For  the more general $\alpha^2\Omega$ dynamo, an extra term $-2R_\alpha (1+\alpha_m)B_r/\pi$ would appear on the right hand side in (\ref{rde2}). This term is negligible compared to the term $R_\omega B_r$, since $|R_\alpha/R_\omega|\sim y^2Ro^{3\gamma}/15\ll1$ using (\ref{dmslp}), and $|R_\alpha/R_\omega|\sim \ty^2/15\ll1$ using (\ref{dmslp'}), for all values of interest of $Ro$.}
\begin{align}
&\partial_tB_r=-\frac{2}{\pi}R_\alpha(1+\alpha_m) B_\phi-\left(R_U+\frac{\pi^2}{4}\right)B_r\label{rde1}\\
&\partial_tB_\phi=R_\omega B_r-\left(R_U+\frac{\pi^2}{4}\right)B_\phi\label{rde2}\\
&\partial_t\alpha_m=-R_U\alpha_m-\frac{\beta_d}{\beta_t}\frac{\pi}{2}\alpha_m-C\left[(1+\alpha_m)(B_r^2+B_\phi^2)\right.\notag\\
&\left.+\frac{3}{8}\sqrt{-\pi(1+\alpha_m)R_\omega\over R_\alpha}B_rB_\phi+\frac{\alpha_m}{R_m}\right]
+\lambda_V\frac{R_\omega}{R_\alpha}(B_r^2-B_\phi^2)\label{rde3}
\end{align}
%EB5 changed eta to beta in the diffusive ratio
where $B_r$ and $B_\phi$ are respectively the radial and azimuthal components of the total magnetic field. The $z$ component is assumed to be much less than these two and is neglected.
%EB4 probably we should add the diffusive flux term and mention it since you nicely considered it... should add references also, including the refs
%HZ4

The second term on the RHS in (\ref{rde3}) governs the effect of diffusive fluxes $\beta_d\nabla^2\alpha_m$ \citep{Brandenburg2009,mitra2010,hubbard2010} where $\beta_d$ is the diffusion coefficient.  For most of the discussion of the solutions in  Sec. \ref{sec4},
we take $\beta_d=0$ (the case of \cite{sur2007})
except for  Sec. \ref{sec4.5} where we adopt $\beta_d/\beta_t=1$ in a model using the radial coordinate $r$ as a free parameter and find that this diffusive helicity flux term
raises the magnetic the saturated magnetic energy  as it exceeds the wind flux term $R_U$ for the fiducial parameters chosen, over much of the disk.
The last term in Eqn. (\ref{rde3}) is the Vishniac-Cho flux \citep{Vishniac2001} with
dimensionless coefficient $\lambda_V$.
We find that that, in accordance with
 \cite{sur2007} that this flux has an influence only after the field already grows
 substantially, and has its strongest influence at low Rossby numbers.
 Even then, the buoyancy flux tempers the influence of the Vishniac-Cho flux. In the solutions
 presented in the sections below, we  focus primarily the case of $\lambda_V=0$.
%EB6 see above:  can you add the vishniac cho flux to 37 but with a coefficient that we can set to zero

%EB5 tweaked above--do you agree with my last qualification "for the fiducial..."?
%HZ5 in fig.8 the value of e_B is almost raised by one order of magnitude.. and i'm not sure if this can be called "minor"..
%EB6 edited it further

%To capture what fraction of the total energy is turned into magnetic energy more clearly,
The magnetic fields are normalized by the equipartition field strength $B_{eq}=\sqrt{4\pi\rho}v$, so that $B_r=v_{A,r}/v$ and so on with $v_A$ the Alfv\'en speed. Note that $B_{eq}$ is a function of $v$ and thus varies with
%HZ 12/14
both the eddy turnover time and the galactic rotation speed.
%HZ 12/14 above
We  normalize the time by the diffusion time scale $h^2/\beta_t$ which again depends on the Rossby number. The dimensionless parameters in the above dynamo equations are
\beq
R_\alpha=\frac{\alpha_0h}{\beta_t},\ R_U=\frac{Uh}{\beta_t},\ R_\omega=\frac{h^2q\Omega}{\beta_t},\ C=2\left(\frac{h}{l}\right)^2,
\eeq
where $U$ is the buoyancy speed in $z$ direction containing both a convective flow part $U_0$ and a magnetic buoyancy part $U_B$; $\alpha$ is normalized by $\alpha_0$; and $l=v\tau_{cor}$
is the correlation length scale of the turbulence.

For the fixed-$\Omega$ case of Sec. \ref{2.1}, substituting (\ref{tc}) and (\ref{v}) into those dimensionless parameters gives
\begin{align}
&R_\alpha=yR_{\alpha0},\ R_U=y^{-1}f^{-5/2}R_{U0}+R_{U_B},\notag\\
&R_{\omega}=y^{-1}f^{-3}qR_{\omega0},\ C=y^{-2}f^{-3}C_0\label{dmslp}
\end{align}
where $R_{U_{B}}$ is the magnetic buoyancy term which will be clarified later.

For the  fixed-$\tau_{ed}$ case of Sec. \ref{2.2}, we use (\ref{tc'}) and $\tilf=v/v_0$ to obtain
\begin{align}
&R_\alpha=\ty \tro^{-1}R_{\alpha0},\ R_U=\ty^{-1}\tilf^{-5/2}R_{U0}+R_{U_B},\notag\\
&R_{\omega}=\ty^{-1}\tilf^{-3}\tro^{-1}qR_{\omega0},\ C=\ty^{-2}\tilf^{-3}C_0.\label{dmslp'}
\end{align}

For the $r$-dependent model in Sec. \ref{secr} we use (\ref{unifiedparameter}) along with $U\propto 1/r$ to get
\begin{align}
&R_\alpha=R_{\alpha0}/2,\ R_U=2R_{U0}/\tr^2F^{5/2},\notag\\
&R_\omega=2R_{\omega0}/\tr^2F^3,\ C=4C_0/\tr^2F^3,\label{dmslpr}
\end{align}
where $\tr=r/r_0$ with $r_0=8\ \text{kpc}$,
and we use the following typical data for our Galaxy to calculate the fiducial values (same as in \cite{sur2007}, for the comparison later):
\begin{align}
&\tau_{ed}=10^{15}\ \text{s},\ v=10\ \text{km/s},\ r\Omega=200\ \text{km/s},\notag\\
&l=0.1\ \text{kpc},\ h=0.5\ \text{kpc},\ U_0=1\ \text{km/s},\notag
\end{align}
which gives (with $q=1$)
\beq\notag
R_{\alpha0}\approx1,\ R_{U0}\approx0.3,\ R_{\omega0}\approx-15,\ C_0\approx50,\ R_m\approx10^5,
\eeq
and the corresponding fiducial Rossby number $Ro_0\approx1$.

The instantaneous dynamo number in the kinematic regime can be defined as the square of the ratio of the coefficients of the amplifying rate terms  $\gamma_g$ and the decay rate terms $\gamma_d$:
%HZ6
%EB reference for dynamo number-In
\beq
D_{ins}\equiv \frac{\gamma_g^2}{\gamma_d^2},
\eeq
with
\beq
\gamma_g^2=\frac{2}{\pi}(1+\alpha_m)R_\alpha |R_\omega|
\eeq
and
\beq
\gamma_d^2=\left(R_U+\frac{\pi^2}{4}\right)^2
\eeq
being respectively, the product of growth and decay terms in (\ref{rde1}) and (\ref{rde2}).
We can define the dynamo growth time,  divided by the diffusion time $\tau_{diff}=h^2/\beta_t$,
as
\beq\label{tdyn}
{\tau_{dyn}\over \tau_{diff}}={1\over \gamma_g-\gamma_d}.
\eeq
%where $\alpha_0\approx0$ is taken at $t=0$.
%EB not sure I understand the latter statement since at t=0 alpha_0 has its kinematic finite value
%We have normalized the time using t.

\begin{figure}
\includegraphics[width=0.9\columnwidth]{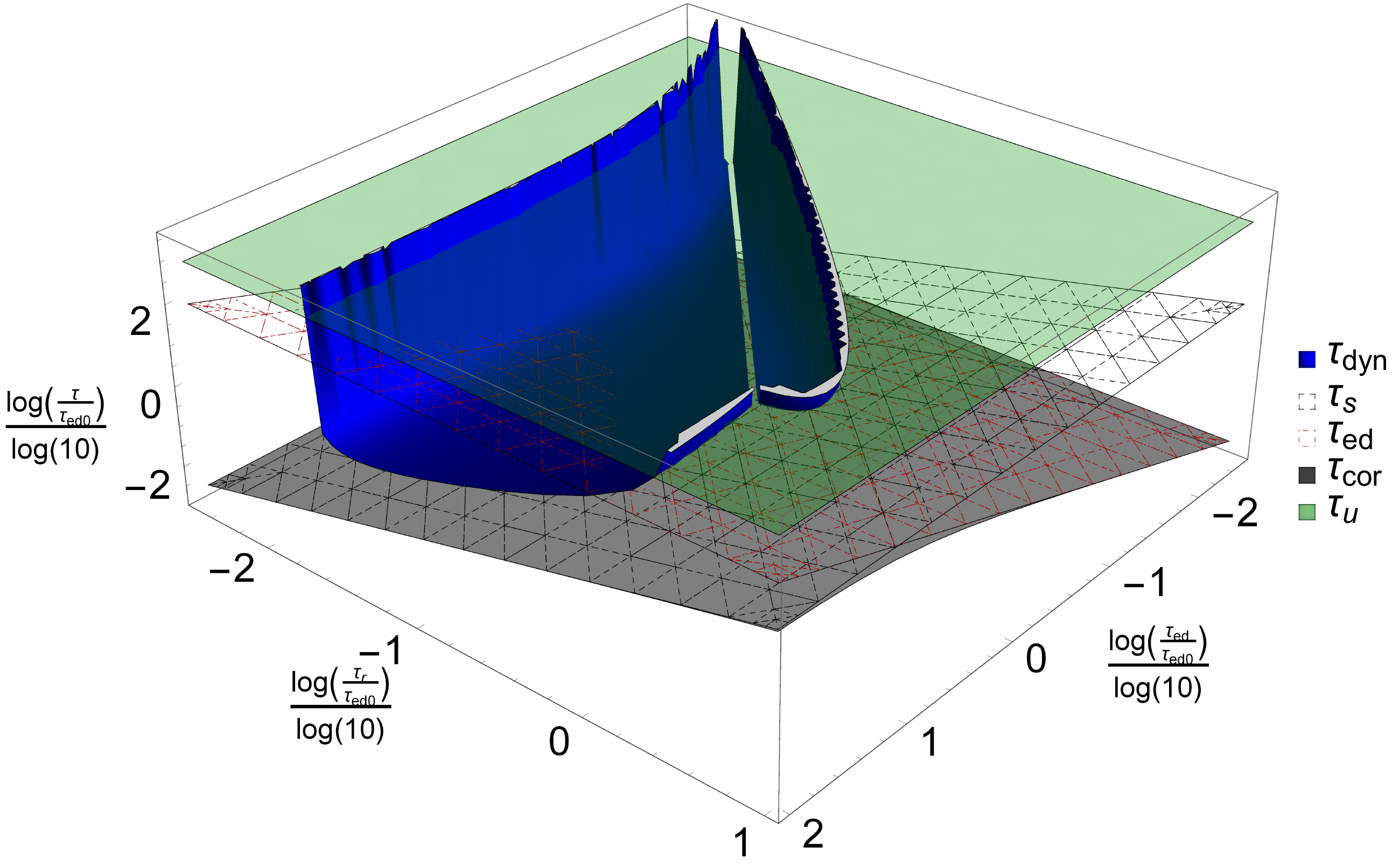}
\includegraphics[width=0.9\columnwidth]{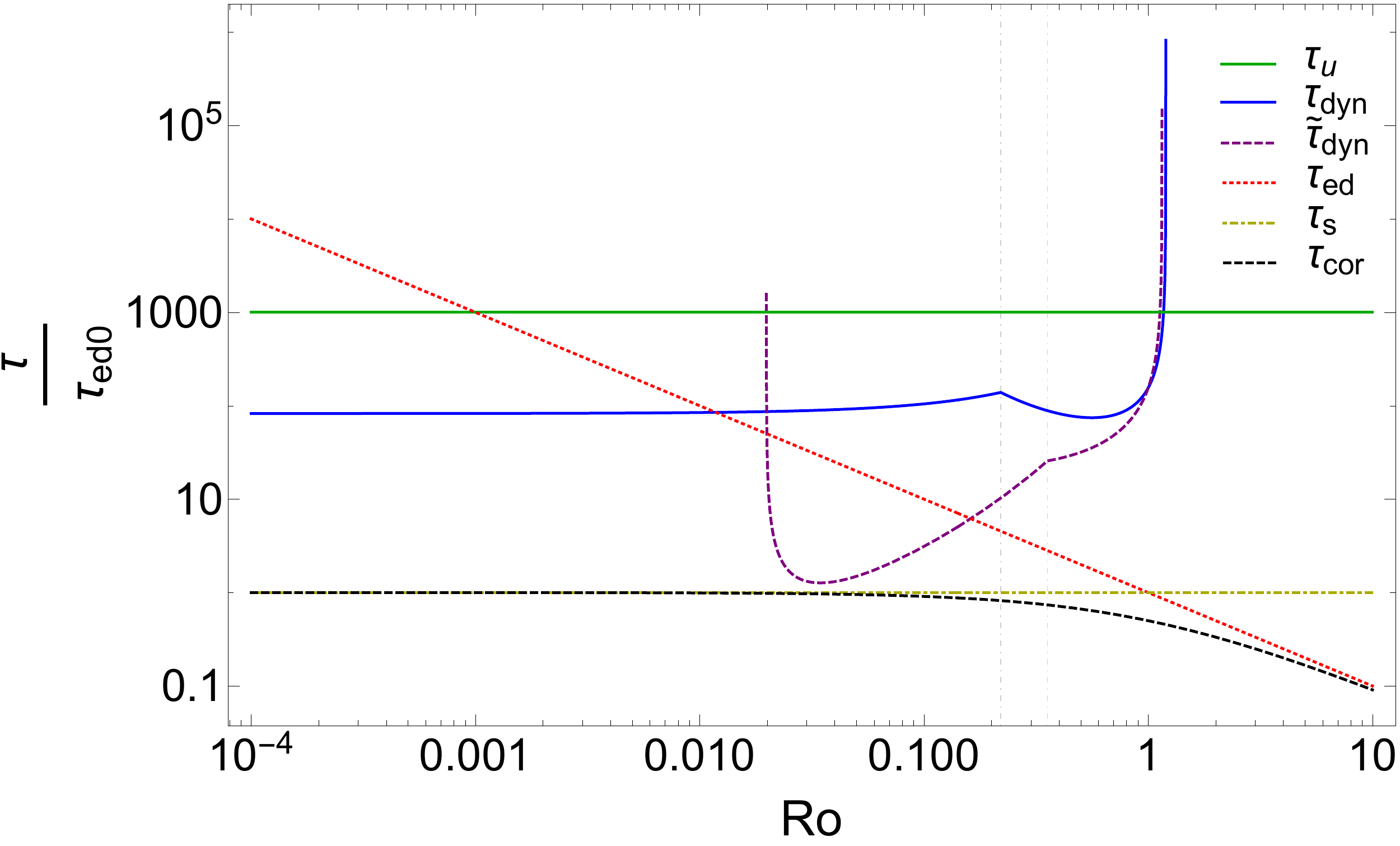}
\vfill
\caption{A Comparison between various time scales. The upper plot is obtained by using (\ref{unifiedf}) which  unifies the results of Sec. \ref{2.1} and Sec. \ref{2.2} into a single
3-D plot.  The dynamo growth time $\tau_{dyn}$ (curved blue surface) is compared with the age of the universe $\tau_u$ (green horizontal flat plane), the eddy turnover time $\tau_{ed}$ (red inclined flat plane), the shear time $\tau_s$ (black inclined flat plane) and the correlation time $\tau_{cor}$ (the lowest gray curved surface). The lines in the  lower 2-D graph shows the quantities at the slice $\tau_r=\tau_{r0}$  corresponding to the  case of Sec. \ref{2.1}  where we vary $\tau_{ed}$ but keep $\Omega$ fixed. The  the $x$-axis in the lower plot is given as the Rossby number.  The purple dashed line shows the growth time, but at a different slice $\tau_{ed}=\tau_{ed0}$, which corresponds to the second case of Sec. \ref{2.2}, where $\tau_{ed}$ is fixed and $\Omega$ varies.
The left and right vertical dot-dashed lines correspond to the transition values of $Ro$ above which supernovae dominate shear turbulence for the fixed $\tau_{r}$ and fixed $\tau_{ed}$ respectively.  }
\label{tdynplot}
\end{figure}

The bottom panel of Fig. \ref{tdynplot} shows $\tau_{dyn}$ (thick blue line) in comparison with the age of the universe $\tau_u\approx 10^3\tau_{ed0}$, $\tau_{ed}$ and $\tau_s$ for our fixed-$\Omega$ case, while the dashed purple line ($\tilde{\tau}_{dyn}$) indicates the dynamo growth time for our  fixed-$\tau_{ed}$ case.   All times in the plot are normalized by $\tau_{diff}$.
The vertical dot-dashed lines  at $Ro=0.22$ and $Ro=0.355$ respectively,  correspond
to the transition values of Eqs. (\ref{f}) and (\ref{f'}) respectively, and
 marking for each of these cases, the transition from shear dominated to   supernova dominated
 turbulent  velocities as $Ro$ increases.   The top panel of Fig. \ref{tdynplot}
 shows the 3-D space that unifies the the cases of Sec. \ref{2.1} and \ref{2.2}
 via Eq. \ref{unifiedf}.

Several interesting  features are evident in the bottom panel of Fig. \ref{tdynplot}.
First, in both two case, for either  the fixed $\Omega$ or fixed $\tau_{ed}$, $\tau_{dyn}^{-1}\rightarrow0$ when $Ro$ approaches $\sim1.2$. As a consequence, for $Ro\gtrsim 1.2$, the initial growth of the magnetic field will be too slow
to produce a significant large scale field.
%EB if t_dyn-> infinity that technically means neglible growth and  in a steady state t_dyn =0 ...so how does that connect with the interpretation above?
Second, $\tau_{dyn}$ becomes independent of  $\tau_{ed}$ when $Ro\leq0.1$ for fixed $\Omega$ (blue curve),
because of the  completely dominance of the shear as a supplier of turbulence.
%(shear driven dynamo?)
In contrast, for the case of  fixed $\tau_{ed}$ (dashed purple line), the growth time
blows up for  $Ro\lesssim0.02$ highlighting that  field growth   becomes insignificant
at these values in this case.  The top  panel of Fig. \ref{tdynplot} shows how these two different cases are mutually compatible in  3-D.
The  solutions  further demonstrating these points will be discussed  in the next section.

For the dynamo to have a significant influence on the large scale field, its  growth time  must be less than the age of the universe $\tau_u$. The associated condition $\tau_{dyn}\le \tau_u$ leads to  an upper bound on  $Ro$ above which the dynamo solution cannot produce significant observable large scale fields .
%finds its maximum or minimum when
In addition,  we impose a lower bound on $Ro$ for fixed $\tau_{ed}$ by  the condition $\Omega_{\text{max}}r=c/10$ with $c$ the speed of light, simply so that we focus on the
the cases where the rotation speed is non-relativistic. Combining these two constraints.
we can  express  the physically meaningful range  as
%EB logarithmic below
%HZ added
\beq
Ro \le 1.171 \ \ (\log Ro\le 0.069)
\eeq
for fixed $\Omega$, and
\beq
0.024 \le \tro \le 1.161 \ \ (-1.62 \le \log \tro \le 0.065)
\eeq
for fixed $\tau_{ed}$.

%\begin{align}
%%=&\left.\frac{1}{\Omega\tau_{ed}}\right|_{\text{max}\{\tau_{dyn}\}=\tau_{u}\text{\ or\ }\Omega r=c/10}\notag\\
%&\left\{\begin{aligned}
%&\ \ Ro \le 1.171 \ \ (\log Ro\le 0.069)\ \ &\text{fixed $\Omega$}\\
%&0.024 \le \tro \le 1.161 \ \ (-1.62 \le \log \tro \le 0.065)\ \ &\text{fixed $\tau_{ed}$}
%\end{aligned}\right\}.
%\end{align}

\section{Solutions}\label{sec4}
For the first 3 subsections below,  we  focus on the fixed-$\Omega$ case,
before addressing a few important features of the  fixed-$\tau_{ed}$ case in the penultimate
subsection. In the last subsection we consider solutions for the case when $\Omega$ and $\tau_{ed}$ mutually depend on $r$.

\subsection{Steady-State Without Magnetic Buoyancy}
We first consider the case without magnetic buoyancy.
 By solving  Eqs. (\ref{rde1})-(\ref{rde3}) for a steady state ($\partial_t=0$) we obtain the darkest blue dotted line in Fig. \ref{totalfield}. The $y$ axis, representing the magnetic field strength, is scaled with the equipartition field strength $B_{eq}^2=4\pi\rho v^2$ which  depends on $Ro$.
 %Therefore, the curves always reflect the relative magnetic energy compared to turbulent energy at each instant $Ro$.
  To the left of the the vertical dot-dashed line at $Ro=0.22$
    the turbulence is mostly driven by shear and right to  by SNe. The cusp
    irregularity at $Ro=0.22$ occurs because of our piecewise-defined (\ref{f}), which in principle can be removed by rigourously solving $f$, but not essential for the  level of detail explored here.
  We used (\ref{f}),
  %a simpler form of $f$,
  which  is sufficient to
  capture the asymptotic behavior for large and small $Ro$.
   The darkest blue dotted-line solution includes  the influence from differential rotation of both $\tau_{cor}$ and $v$ and can be compared  with the top dotted line, obtained by taking for the full range of $Ro$
 \beq\label{pureshear}
f=(Ro^4 y)^{1/5};\ y=\text{min}\{Ro^{-1},Ro^{-1/2}\},
\eeq
which is the expression for $f$ that neglects the effect of shear in the turbulence and correlation time (though shear is still maintained for the $\Omega$ effect in the ${\overline B}_\phi$ equation).

%EB commented out paragraph below because helical turbulence shouldnt require shear --but then one has alpha^ 2 dynamo  which we are not considering here.  also the q=22.7 is very large.
%The distinct influence  of differential rotation that manifests from Eq. (\ref{f})  can be checked by setting $Ro=1$ and recovering the factor $q$ in the equations, and using it as a free parameter.
%This procedure shows that the dynamo is essentially quenched for both small ($q \lesssim0.73$) and large ($q\gtrsim22.7$). The former case corresponds to weak shear, where little helical turbulence can be produced and thus there is no mechanism for the toroidal field to amplify the poloidal one.
 %On the other hand, in the latter case the shear is too strong that not enough helicity is carried by eddy before new eddies are produced and diffuse the old ones, causing a quenching in the $\alpha$ effect.

%Another thing to notice is our arbitrariness of the choice of function $y$. If we write $y$ in the form (note $s$ is a function of $Ro$)
%\beq
%y=\frac{2\pi s}{1+2\pi s Ro+2[2 \pi s Ro]^\mu},
%\eeq
%actually we have the freedom to choose any $\mu$ in the range $(0,1)$, without violating constraints (\ref{cst1}) and (\ref{cst2}). It can be verified (not demonstrated here) that different choices of $\mu$ do not affect the characteristic of our result, only causing a translation of the curve along the $Ro$ axis.

In Fig. \ref{bphibr} we show how different components of the magnetic fields depends on the Rossby number. We define the pitch angle  by
%HZ6
%EB need ref here, maybe one referred to in Chamandy
\beq
p\equiv \arctan\frac{B_r}{B_\phi}=\arctan\frac{R_\omega}{R_U+\pi^2/4}
\eeq
where we have used (\ref{rde2}) for the last equality. The magnitude of $p$ decreases with  decreasing $Ro$ when the turbulent energy is mostly provided by SN  (region to the right to the vertical line), in agreement with the numerical solution in \cite{chamandy2016statistical} (see their Fig. 2, where they used the Coriolis number $Co=1/Ro$). As expected, the pitch angle goes to a constant as $Ro\rightarrow0$, since without SN, the turbulent energy and the correlation time depend only on the rotation profile. Then $Ro$ drops out of the equations and the dynamo saturates to a state purely driven by shear at fixed $q$.  The smallness of the pitch angle is consistent with  the basic observation that galactic magnetic fields are predominantly azimuthal \cite{BeckWielebinski2013}.
%HZ6
%EB ref to observations

\subsection{Role of Magnetic Buoyancy vs. Outflow and Diffusive Flux}
%EB4 added diffusive flux here--worth adding a sentence or two somewhere at the end of this section to comment on the diffusive flux, even if we only consider it  explicitly for the r dependent solution case later in section 4.5
We now investigate the inclusion of  magnetic buoyancy.
Although  \cite{foglizzo1994} suggest  that differential rotation will stabilize the Parker mode, we neglect this effect in our rough calculations here.
We use the buoyancy  speed as  calculated in \cite{parker1979}. For a weak magnetic field of sub-equipartiation (with the turbulence) strength,  $U_B\approx v_A^2/v=v(B_r^2+B_\phi^2)$.
For a magnetic field comparable to  equipartition strength, $U_B\approx v_A=v\sqrt{B_r^2+B_\phi^2}$. The field-related buoyancy  coefficient (assuming $|B_\phi|\gg |B_r|$) is then
\beq\label{mb}
R_{U_B}=\frac{U_Bh}{v^2\tau_{cor}}=\left(\frac{C_0}{2}\right)^{1/2}y^{-1}f^{-3/2}\text{min}\{B_\phi,B_\phi^2\}
\eeq
for fixed $\Omega$.
(For the case of fixed $\tau_{ed}$ we would just replace $y$ by $\ty$ and $f$ by $\tilf$ in the above expression.)
% A self-consistent solution is given in

\begin{figure}
{\includegraphics[width=0.9\columnwidth]{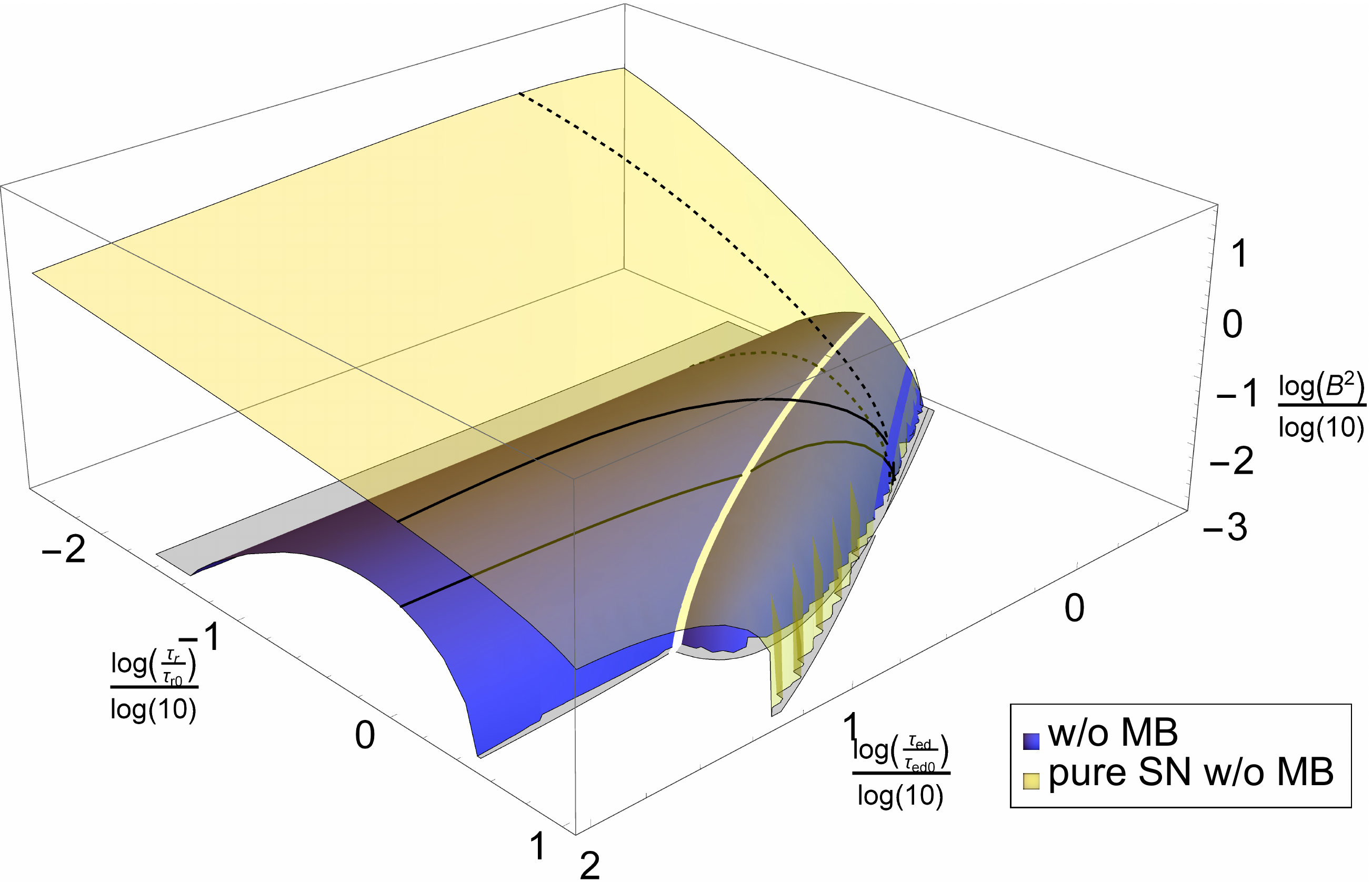}
\includegraphics[width=0.9\columnwidth]{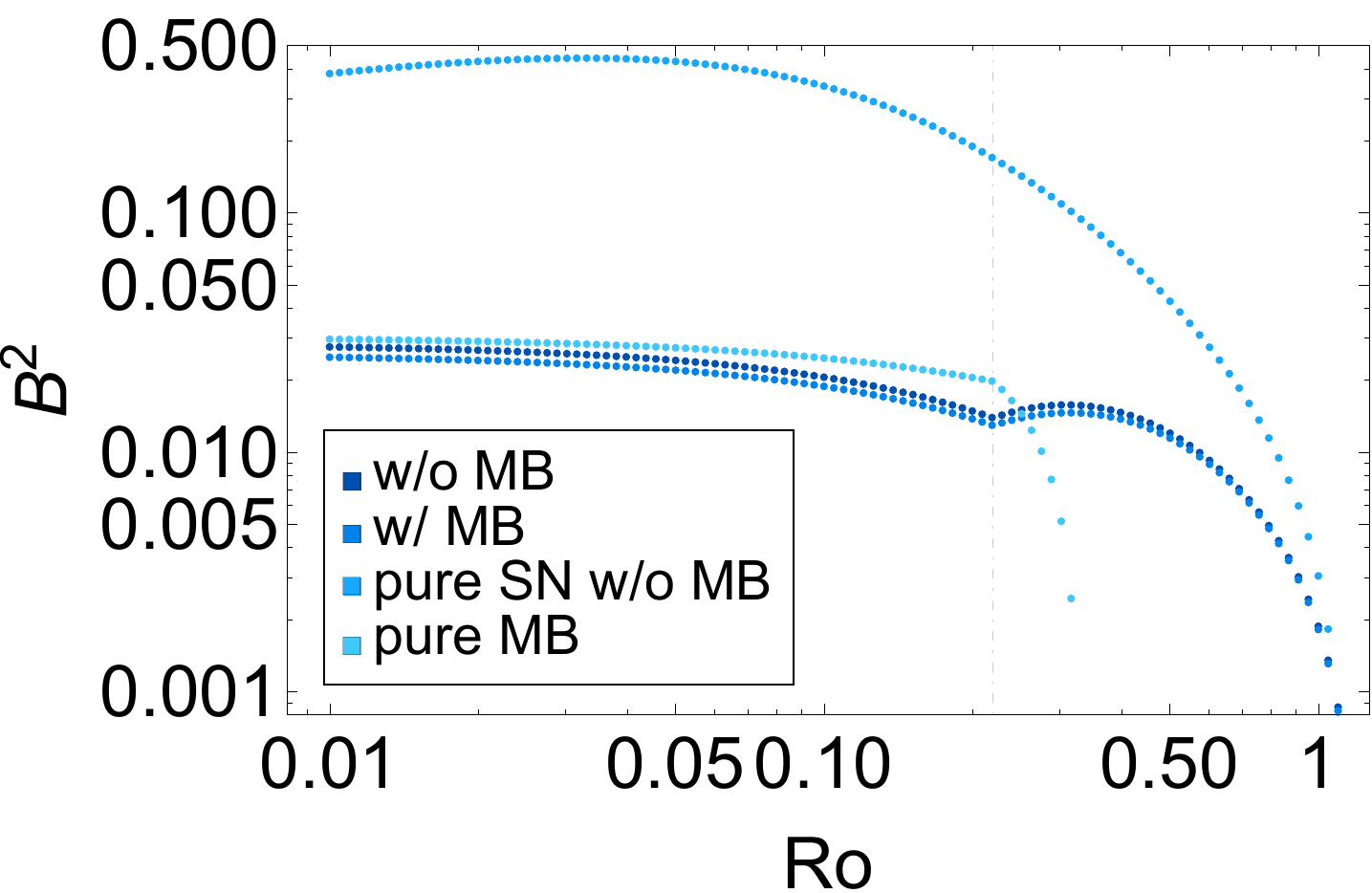}}
\caption{
Magnetic energy $B^2=B_\phi^2+B_r^2$ as a solution of Eqs. (\ref{rde1})-(\ref{rde3}).
(a) Full 2-D solution surface showing the dependence on both $\tau_r$ and $\tau_{ed}$ without magnetic buoyancy (blue surface) compared to the 2-D solution surface in the absence of the influence of shear on the turbulence. The solid (dashed) black curves are obtained by taking a slice at the fiducial value $\tau_r/\tau_{r0}=1$ ($\tau_{ed}/\tau_{ed0}=1$), corresponding to the case with fixed $\Omega=\Omega_0$ ($\tau_{ed}=\tau_{ed0}$) and varying $\tau_{ed}$ ($\Omega$).
(b) The  1-D solution  for fixed $\Omega$ but varying $\tau_{ed}$, corresponding to the solid red curves in (a).
The two darkest dotted curves represent the results with and without magnetic buoyancy, respectively. The competition between enhanced large scale growth from ejection of buoyant ejectio of small scale magnetic helicity with the buoyant loss of large scale fields
 can be assessed by comparing the two curves. The top dotted curve is  obtained by neglecting the contribution from shear in both the turbulent energy and the correlation time. Dynamos supported purely by shear and magnetic buoyancy become possible for small $Ro$, indicated by the lowest dotted curve.}
 %EB2 edited caption above
 \label{totalfield}
\end{figure}

MB extracts  small scale magnetic helicity but also large scale fields.  As a consequence, there is a competition between the loss of large scale field and benefit to  amplification from small scale magnetic helicity removal.
The bottom dotted line of Fig. \ref{totalfield} shows the solution for the fixed $\Omega$ case.
Here the presence of MB lowers the overall field strength compared to the case
when $R_{U_B}=0$.
For fixed $\Omega$, we also note the possibility of  dynamo purely supported by only magnetic buoyancy, where $U_0=R_{U_0}=0$. This solution is represented by the lightest blue curve  in Fig. \ref{totalfield}.

\begin{figure}
\includegraphics[width=0.9\columnwidth]{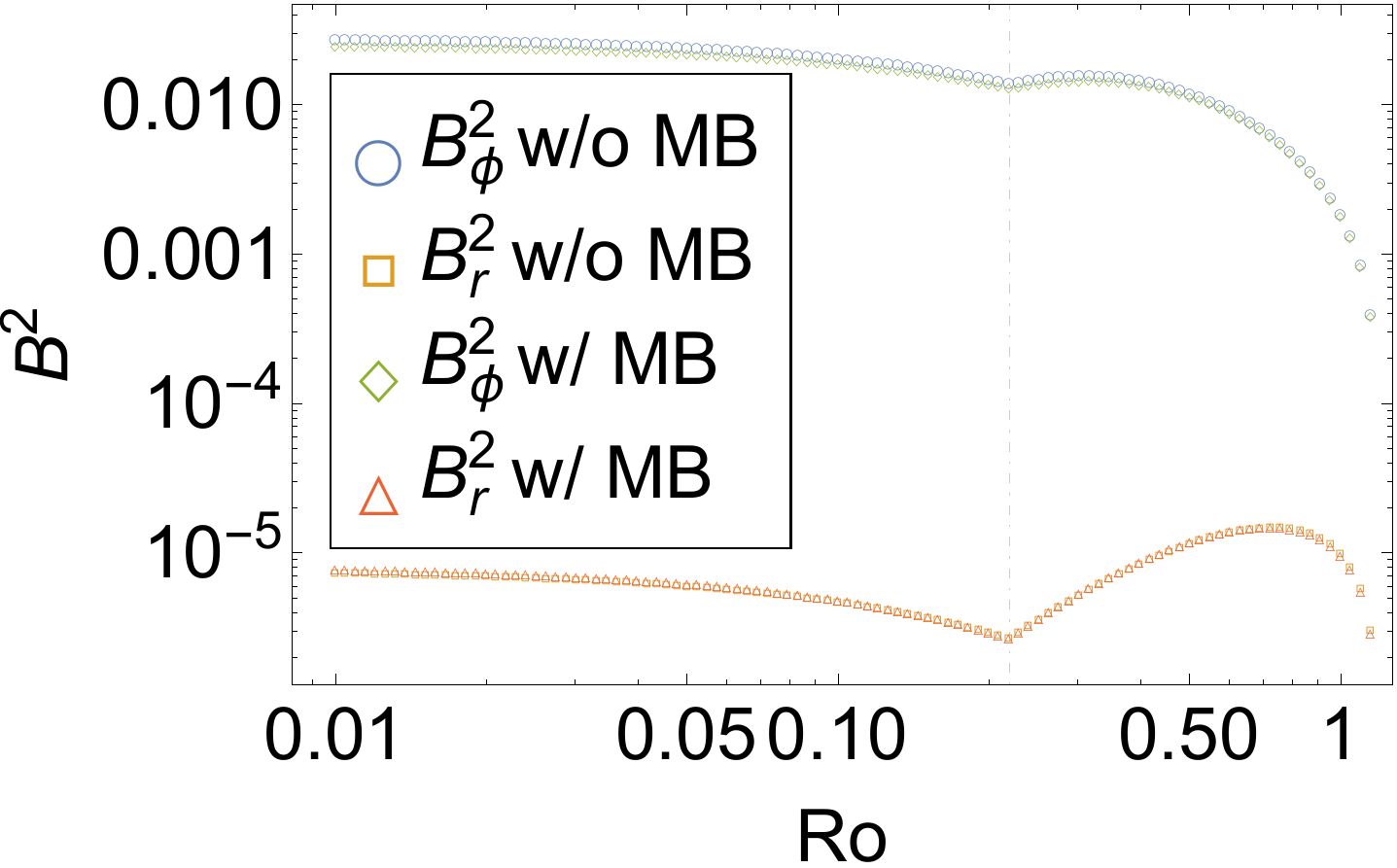}
\vfill
\includegraphics[width=0.9\columnwidth]{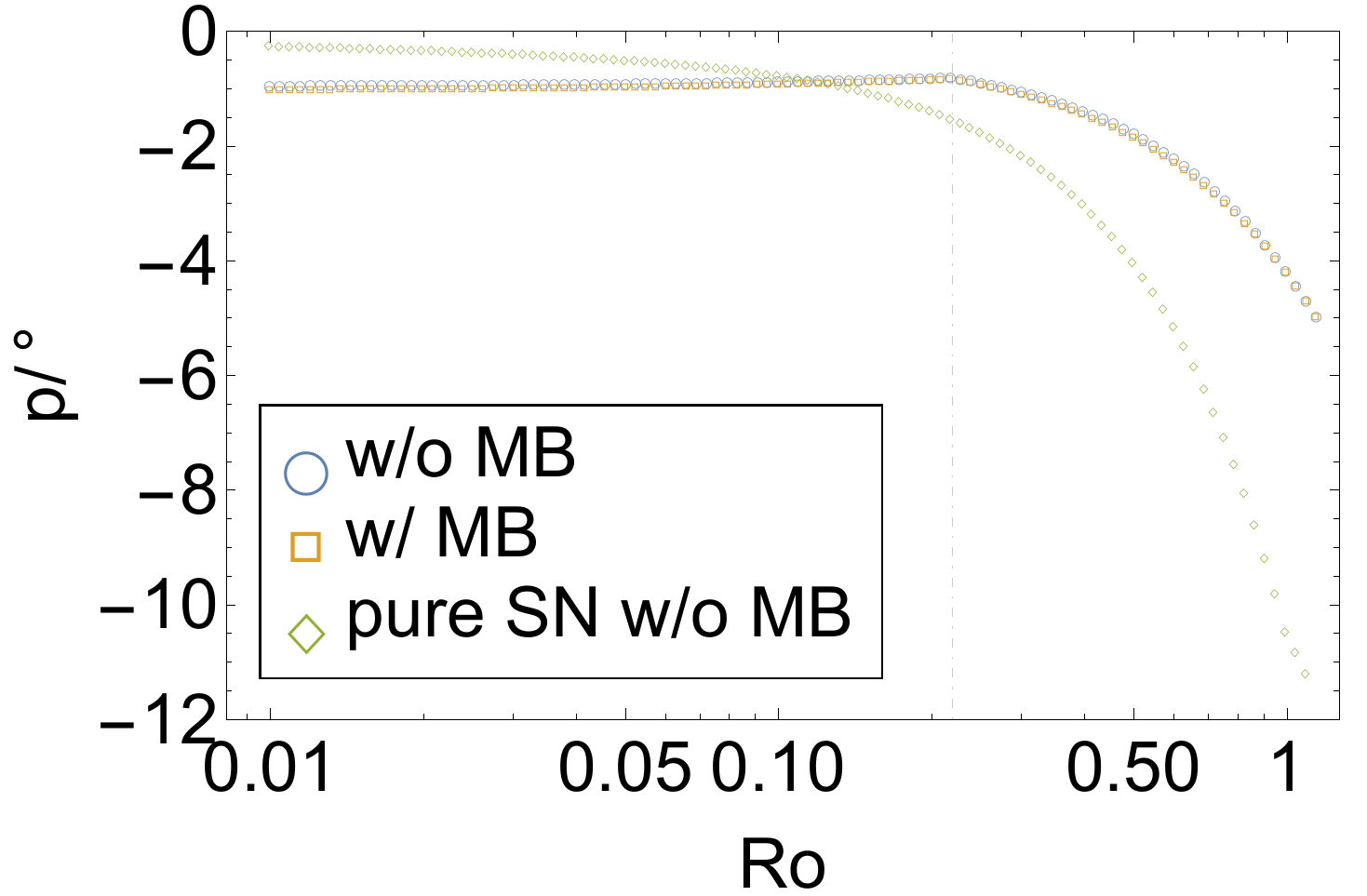}
\caption{Upper: The saturated values of toroidal ($B_\phi$) and  poloidal ($B_r$) components of the field are shown, for cases with and without MB. The toroidal fields are mostly suppressed, while the poloidal fields are amplified when MB is included. Lower: The pitch angle $p=\arctan({B_r}/{B_\phi})$ is presented, showing the relative strength of the two components of the magnetic field. The increase in $|p|$ when MB  is included results from a greater loss in the toroidal field.}
%EB increase in Ro decreases the ratio of Br/B_phi above Ro=0.2 hmm...lets discuss
\label{bphibr}
\end{figure}

The curves represented by green diamonds and red triangles of Fig. \ref{bphibr} show the different behaviors of the toroidal and poloidal magnetic fields. The growth of toroidal field ($B_\phi$, blue circles and green diamonds) is suppressed by MB, whereas the poloidal field ($B_r$, yellow squares and red triangles) is amplified by MB. This is understandable by noting the competing roles of MB  mentioned above, and the fact that in Eqs. (\ref{rde1}) and (\ref{rde2}), MB is more significant for the toroidal field loss because $|B_\phi|\gg |B_r|$.

%HZ4
The importance of the diffusive helicity flux (second term on the right of Eq. (\ref{rde3}))
can be assessed by its separate ratios to the  wind term (first term on the right of Eq. (\ref{rde3}) and the MB (third term on the right of Eq. (\ref{rde3})).
For $ \beta_t=\beta_d$ these are respectively
%EB5 I referred to the specific equation where the ratio is taken..but you might also want to add what other equation is used in this relation below.
\beq\label{difftowind}
\frac{diff}{wind}=\frac{\pi}{2R_{U_0}}yf^{5/2}
\eeq
and
\beq
\frac{diff}{MB}=\left(\frac{\pi^2}{2C_0}\right)^{1/2}yf^{3/2}\text{min}\{B_\phi,B_\phi^2\}
\eeq
in the case of fixed $\Omega$. Since both ratios are smaller than 1 when $Ro<1$, keeping or neglecting the diffusive helicity flux term will not change the results significantly.

The pitch angle profile under the influence of MB is shown in Fig. \ref{bphibr}. this curve explicitly reveals that MB more strongly suppresses azimuthal fields.

\begin{figure}
{\includegraphics[width=0.9\columnwidth]{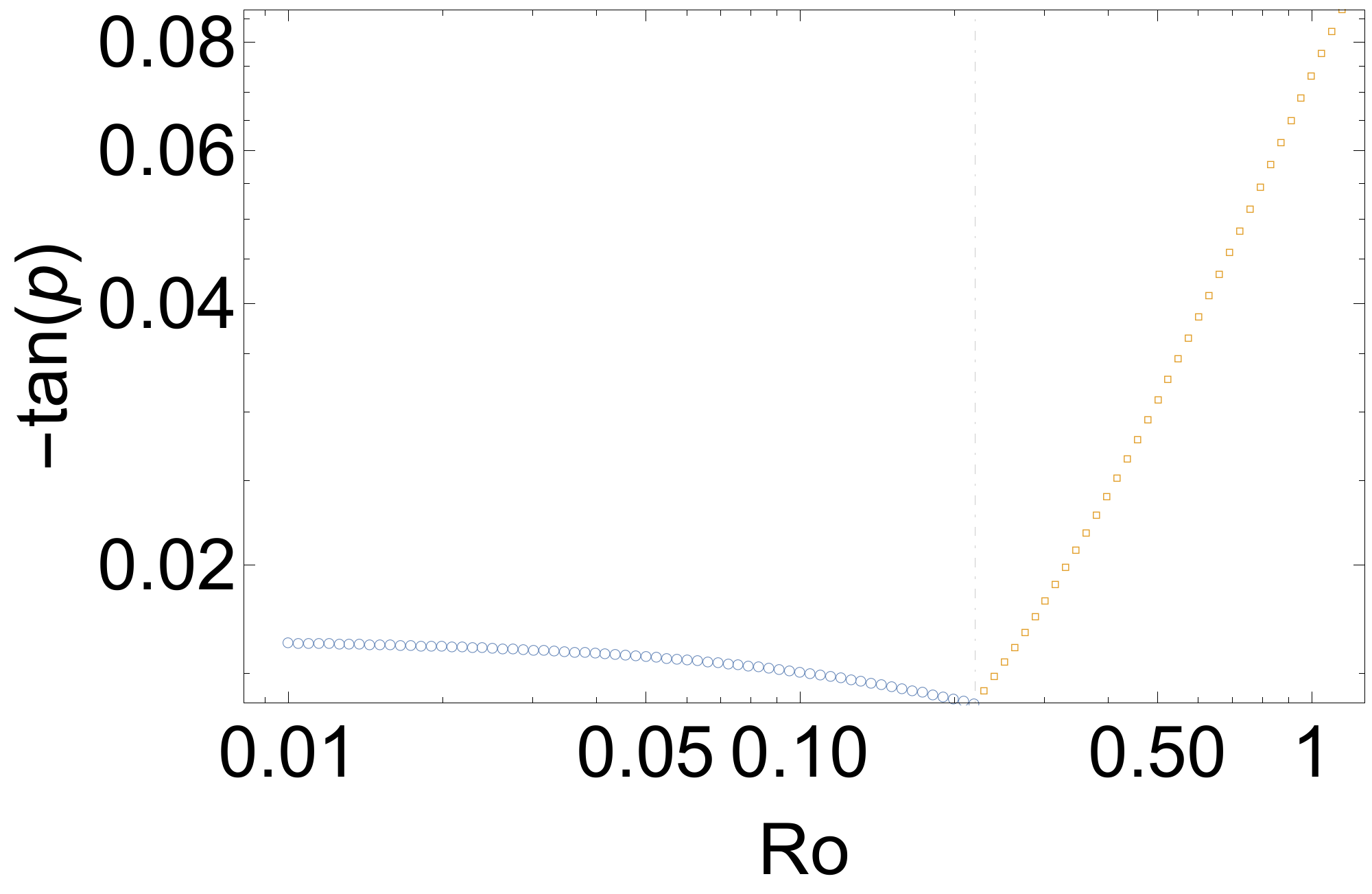}}
 \caption{The tangent of the pitch angle $p$ as a function of $Ro$ in a fixed-$\Omega$ model. In the right half region a power law $-\tan p\sim Ro^{1.13}$ is found using the red data point.}
 \label{tanp}
\end{figure}

%EB4  perhaps we should we add plot for  adding the prediction for tan p vs. Ro for this case, i added this  text below based on your email note for you to edit further...
%HZ4
For this model, we can predict  the tangent of the pitch angle as a function of $Ro$. The result is shown in Fig. \ref{tanp} where we compare our numerical prediction with that of \cite{chamandy2016statistical} (who found $\tan p\sim\tau_c (v/h)^2/(q \Omega)$).
The red part shows a power law  $\ln(-\tan p)=1.13 \ln(Ro)+constant$.  The limited data in \cite{VanEck2015} from their Fig. 8 suggests a slope of 0.4-0.5,
if we assume that the surface SFR density $\propto$ surface SNR density $\propto1/\tau_{ed}\propto Ro$.  This is closer to the predicted value of  \cite{Chamandy2016} than ours, but more data and work are ultimately needed to pin down the tightness of these trends and predictions.

\subsection{Time-dependent solutions}
\begin{figure}
{\includegraphics[width=0.9\columnwidth]{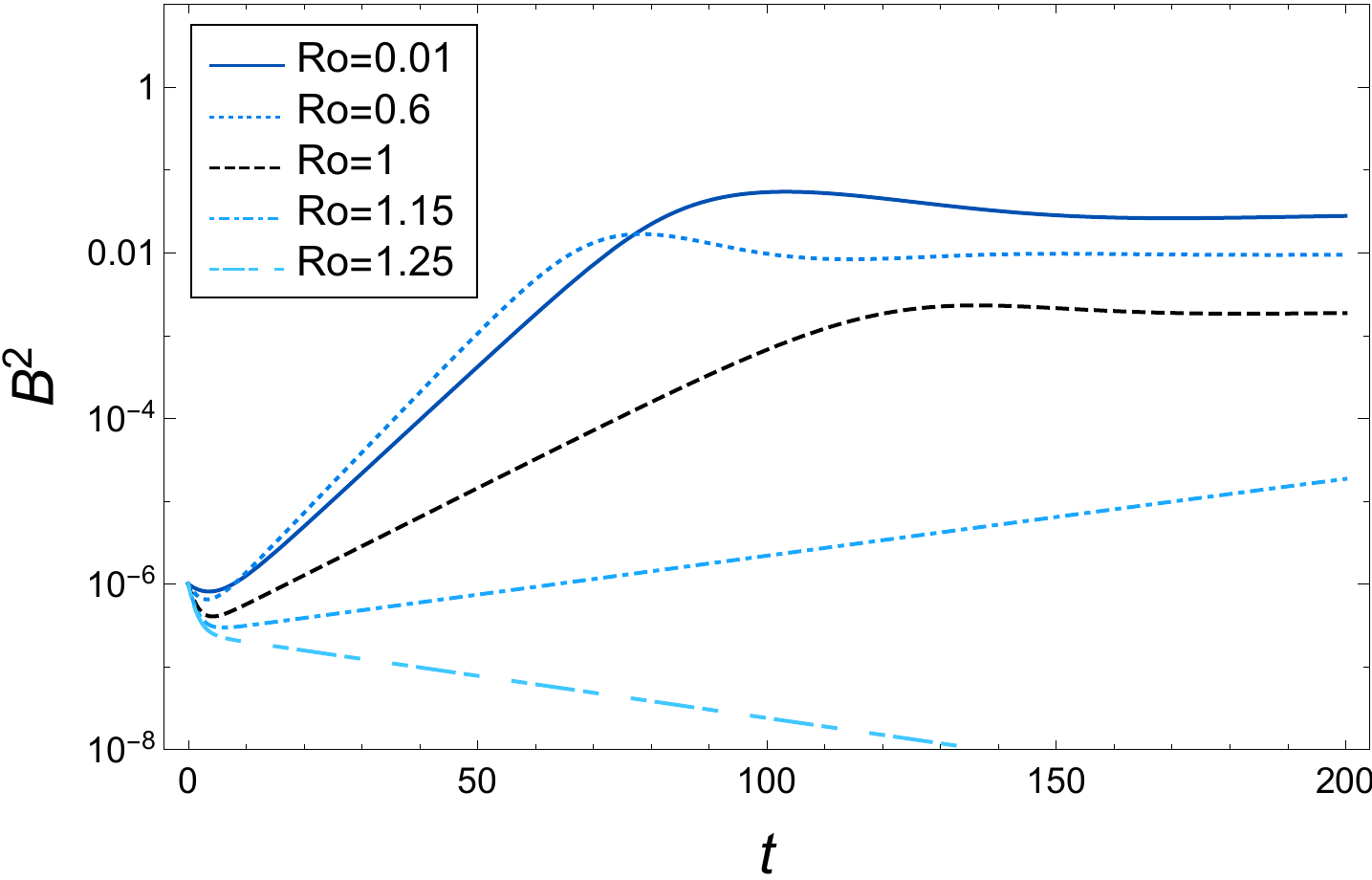}}
 \caption{Time evolution of the total fields for different values of $Ro$ for the fixed $\Omega$ case. Time is normalized by the rotation period $\tau_r$, and $B^2$ is normalized by an $Ro$-dependent equipartition strength in which $v$ is determined by (\ref{v}) and (\ref{f}). Different curves correspond to different values of $Ro$. A non-trivial steady state does not exist for large $Ro$.
 This plot can be viewed as a generalization of  Fig. 2 of  Sur  et al. (2007) to include the effects of shear described
 in the text.}
 \label{timeevol}
\end{figure}

We now compare the time evolution of magnetic fields from the dynamo solutions
for different values of $Ro$ in Fig. \ref{timeevol}. The time is normalized by the constant $\tau_r=2\pi/\Omega$, and the magnetic fields are normalized by the ($Ro$-dependent) equipartition field strength.

The two lower curves show the transition from decaying solutions to those with an asymptotic sustenance of a steady-state as $Ro$ is dialed below $\sim1.2$.
 As $Ro$ is deceased downward from 1.25, the dynamo growth time deceases. The growth time reaches a minimum (the dotted curve, $Ro=0.6$) and then increases,  finally saturating (the solid curve), in agreement with Fig. \ref{tdynplot}. The dashed black curve indicates the fiducial point $Ro=1$.

\subsection{Fixing $\tau_{ed}$ and changing $\Omega$}

\begin{figure}
{\includegraphics[width=0.9\columnwidth]{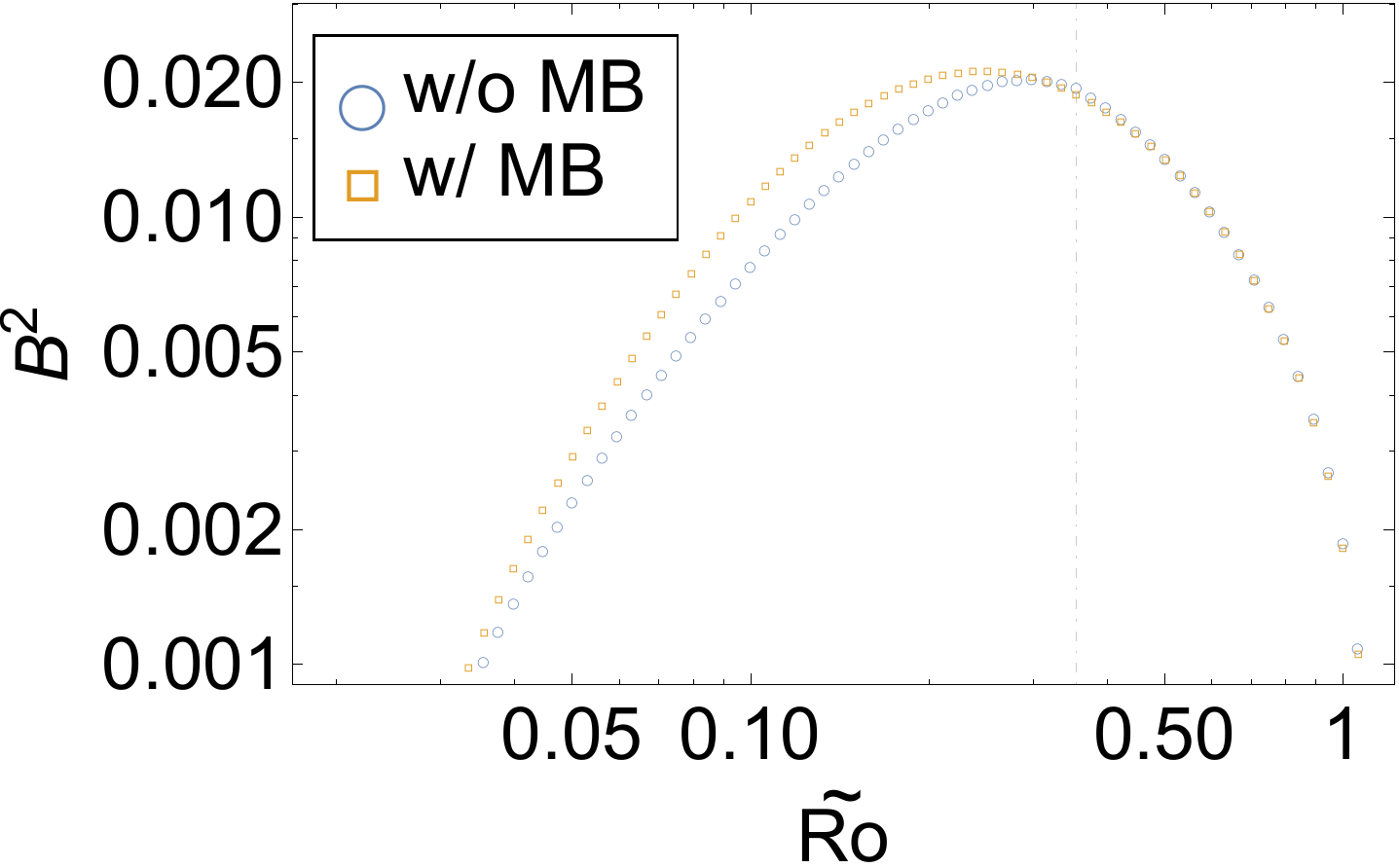}}
 \caption{The solution of the total field for the fixed $\tau_{ed}$ case but allowing $\tro$ to vary by varying  $\Omega$. The plot uses the relation (\ref{y'}). We show solutions  for cases with and without magnetic buoyancy. This plot shows that at both low and high $\tro$ (fast and slow rotation)
 limits, no steady-state solution exists.}
 %EB it occurs to me that someone may ask, no steady state but maybe still growth and cycle period which would be OK on galactic times. Someplace in the text we might address this
 \label{totalfieldo}
\end{figure}

Fig. \ref{totalfieldo} shows the dynamo solutions using the relation (\ref{y'}) and the corresponding non-dimensional parameters in the dynamo equations for the case
of fixed $\tau_{ed}$ and varying $\Omega$.  The vertical dot-dashed line marks the
 transition value $\tro=0.355$ between shear-dominated and SN dominated turbulence. The maximum steady-state field strength  $\sim0.02 B_{eq}^2$ occurs at intermediate $\tro\sim0.2$, and  decays with $\tro$ for both lower and higher $\tro$. This  contrasts the saturated steady states
 of Figs. \ref{totalfield} and \ref{bphibr}  for small $Ro$ where we fixed $\Omega$ and allowed $\tau_{ed}$ to vary.

\subsection{Dynamo solutions as function of radius when  SN rate depends on  $\Omega$}\label{sec4.5}
\begin{figure}
{\includegraphics[width=0.9\columnwidth]{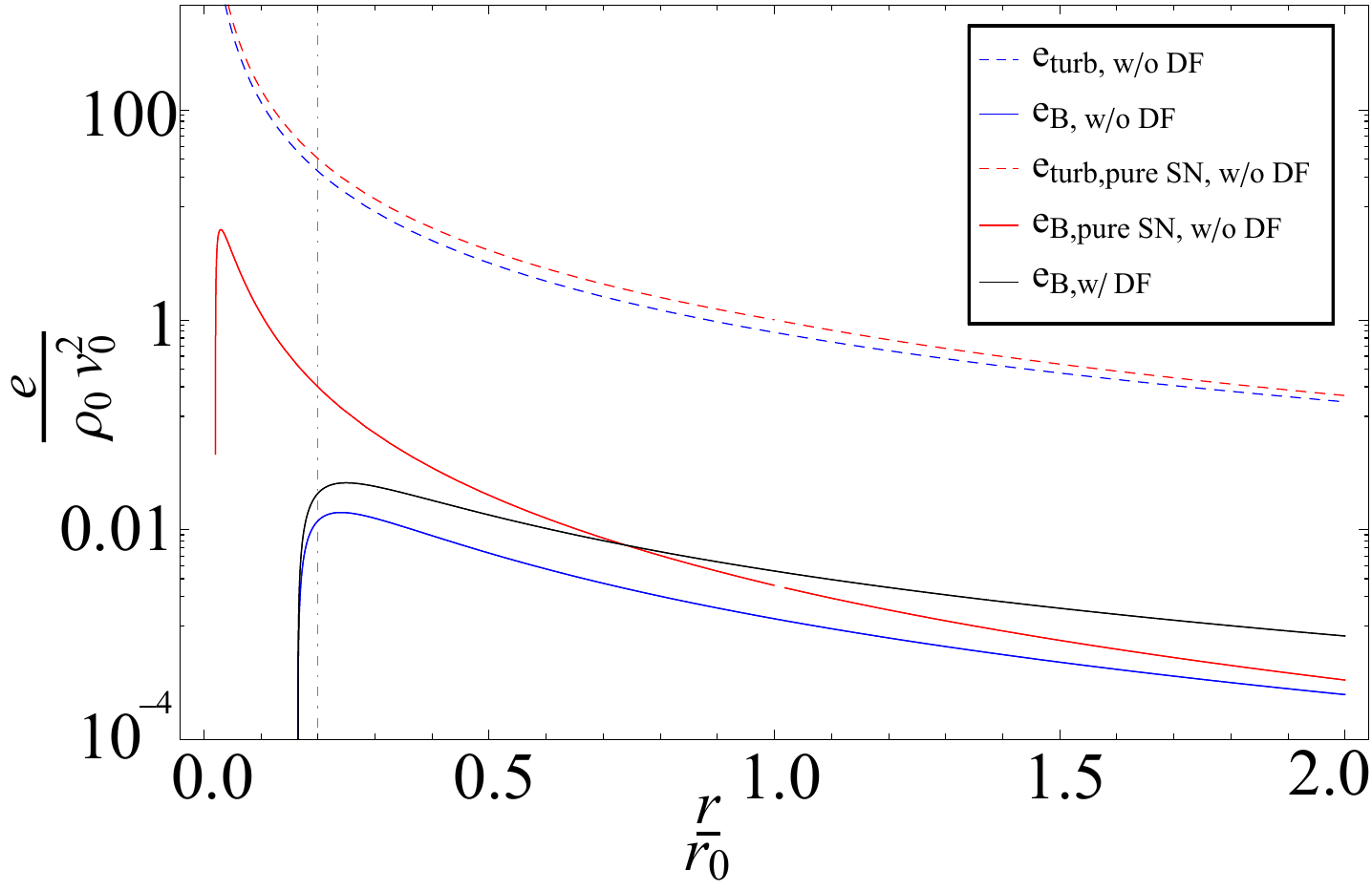}}
 \caption{Turbulent energy density $e_{turb}$ and magnetic energy density $e_B$ as functions of $r$ in a model where both $\tau_{ed}$ and $\Omega$ depend on $r$ only.
 %HZ4
 The vertical dashed line is at $r/r_0=0.2$ to show roughly where the galactic central region is.
 The blue curves consider the effects of both SNe and shear while the red curves only consider the former. Comparing the two cases, we see that the magnetic energy density becomes flat if we take the effect of shear on correlation time and turbulent energy density into consideration.
 %EB4 comment on what the vertical line means
 }

 \label{rmodel}
\end{figure}

Fig. \ref{rmodel} shows the result in using the model discussed in Sec. \ref{secr} and (\ref{dmslpr}). The horizontal axis is normalized by $r_0=8\ \text{kpc}$. Here we define $e_{turb}=\rho v^2=\rho_0 v_0^2 f^2/r$ and $e_B=B^2$ as the turbulent energy density and magnetic energy density, respectively, and show them in blue curves. The ISM mass density is assumed to have the same dependence on $r$ as the galactic surface density, i.e., $\rho\propto 1/r$. Red curves represents the model with (\ref{pureshear}) being used, i.e., it neglects the effect of shear on both correlation time and turbulent energy density. Beyond the galactic central region $r/r_0<0.2$ where a more sophisticated dynamo model is needed, we obtain a nearly flat profile for both turbulent and magnetic energy density in agreement with Fig. 20 of \cite{BeckWielebinski2013}.
%HZ4
%EB4 yes, add the observational refs here and again in conclusions, where i made a note

Curiously if we compare the two dashed lines, i.e., the turbulent energy densities with and without considering the shear, the latter is above the former yet the former one includes energy sources from both SN and shear. This is not surprising if we realize that even though cooperating shear into the model increases the turbulent energy input rate, the correlation time is decreased at the same time, leading to a net effect of lowering $e_{turb}$ (see Eq. (\ref{turbngy})).

%EB4 worth mentioning the case where you tried diffusive flux  here.. maybe we use that case for the plot of figure 7. I pasted your text from your email as a starting point that can be edited: "made the diffusive coefficient=eta_t so the result will be somehow overestimated.  After non-dimensionalization it is an extra term (-Pi/2 alpha_m) on the RHS in (37) in the draft. So indeed one is to compare Ru with Pi/2 to see whether the advection flux or the helicity flux dominates. In fact Ru is comparable to Pi/2 when r<0.3, and Pi/2>>Ru when r>0.3. The result is that the magnetic energy density is raised a little bit by this term and is somehow flattened. It is not a significant change in the curve."}
%HZ4
The black curve of Fig.  \ref{rmodel}  represents the magnetic energy density if we take the diffusive helicity flux term into consideration.  Here the diffusion coefficient $\beta_d$ is assumed to be equal to the turbulent diffusivity $\beta_t$, which may be an overestimate because usually the ratio $\beta_d/\beta_t$ is taken to be $\ll 1$, e.g. in \cite{BCC2009} it is 0.05, and in \cite{mitra2010} a value of $\sim0.3$ is found (albeit at very low $R_M$ compared to what would be appopriate for galaxies).
Using (\ref{ry}), (\ref{rf}) and (\ref{difftowind}), we find that the contribution from the wind term (characterized by $R_U$) is comparable to that from the diffusive term (characterized by $\pi/2$) when $r<0.3$, and $\pi/2\gg R_U$ when $r>0.3$, showing a dominance of this diffusive helicity
%EB5 added diffusive
 flux in almost the whole disk. The  inclusion of $\beta_d$  increases the saturated value of magnetic  energy by nearly an order of magnitude given our fiducial parameter choices.
 %EB5 edited above. the change as you say is kind of significant. On the plot, change HF to DF :)
 %HZ5 changed

%14 times the equipartition field strength $B_{eq}^2=4\pi\rho v^2$ is observed. This is possibly because we have input an increasingly larger amount of energy in the form of galactic rotation into the system as we turning down $Ro$, but the energy scale keeps the same. One way to see this is to compare $B_eq^2$ with the equipartition field strength using a velocity scale $\Omega r$, which yields a ratio $v^2/\Omega^2r^2$ of $10^{-4}Ro^2$. Thus, even for small $Ro$, the field strength is only a small fraction of the available rotation energy. Another possible way to resolve the problem is to argue that $\rho$ should be positively correlated to the rotation speed so that we should have a larger equipartition value for smaller $Ro$.

\section{Conclusions}\label{sec5}

We have generalized  a 2-D  ``no-z'' galactic dynamo model with helicity fluxes to include two effects of differential rotation beyond its role in the $\Omega$-effect which have not been previously combined in galactic dynamo models.  First,  differential rotation provides an additional energy source for  ISM turbulence, beyond
that of  SN.
%which can dominate in a region of a galaxy where the the SN input rate is low and independent of the rotation rate.
Second, differential rotation can shred turbulent eddies,  reducing the  correlation time of the turbulence
\citep{blackman2015}.

We have incorporated these effects and  relaxed the commonly assumed
equality between the correlation time and the  SN driven eddy turnover time.  We show that the effect of shear on the correlation time
 can be important even when  shear does not  dominate the turbulent energy.  For low SN rates and strong shear, both
effects are important.
We  separately studied the influence  of differential rotation on the mean field dynamo solutions as a function of  the SN input rate and the rotation period  when these quantities are taken to be independent and also when they are proportional  to each other. The latter would be  expected from  correlations of the SN rate with the star formation rate and in turn,  the galactic surface density and  rotation rate \citep{prasad2016}.
%EB3 ref above
%With these ingredients we solved the mean field "no z" galactic dynamo equations
%with helicity fluxes.
%We
% strengths and pitch angle in saturated states and time evolution of the fields are presented in Sec. \ref{sec4}.

Our solutions show that  the observable steady-state mean field dynamo field strengths
at low Rossby numbers are significantly lower than those  when the correlation time is independent of shear.
 %than being catastrophically divergent if we have simply adopted $\tau_{cor}=\tau_{ed}$ and ignore the turbulent energy from shear.
 The  model also predicts the  pitch angle of the mean field, a measure of radial to toroidal field magnitude, and a  clean
 quantity to compare with observations \citep{Chamandy2015}.
%more detailed curve of $\tau_{cor}/\tau_{ed}$.
Unlike previous work, we have also included magnetic buoyancy  as a contributor to the helicity fluxes
which becomes most  important when $Ro,\tro<1$.
We find that dynamos for which the helicity fluxes are entirely determined by
buoyancy are possible even in the absence of advective or diffusive fluxes.

%In our galaxy, the role of  first role is usually neglected. However, one could imagine for low supernova rate or strong shear regions, i.e., where the Rossby number is small, turbulence driven by shear becomes dominant. In the present paper we use a physical argument (\ref{turbngy}) to capture this first role, that the total turbulent energy density input rate should be equal to the sum of the contributions from supernova and from shear.  As for the second role, in most literatures the correlation time $\tau_{cor}$ in the dynamo equations is usually identified with the eddy turnover time $\tau_{ed}$.

We also considered a model (Sec. \ref{secr}) where both $\tau_{ed}$ and $\Omega$ are functions of $r$. All dimensionless parameters were then reinterpreted as functions of $r$ only, as in (\ref{dmslpr}). When our model is used in this way to explore radial dependence of quantities within a galaxy, we derived that the  magnetic energy density profile is relatively flat in radius, consistent with observations \citep{BeckWielebinski2013}
%EB4 refes
%EB4 there are a few caveats---e.g. when the Schnmdt -kennicut index is 1,  perhaps  worth mentioning?
and it is a result that serves as a test/consistency check
for the model.
%HZ4
The shape of the curve is sensitive to the Schnmdt-Kennicut index of Sec. \ref{secr}. If we switch it from 1 to 1.4, the radius at which the steady-state magnetic energy density drops to zero will move from $r\sim0.2$ to $r\sim 0.7$.

Earlier prescriptions for galactic dynamos with  $Ro<1$  included only
the effect of $\Omega$ on the reduction of correlation time
 in $\alpha_0$ \citep{rss1988}, and without explicitly including the role
of shear as a source of energy for the turbulence. We showed herein that shear
causes a further reduction in the correlation time not captured by the previous treatments
%but not shear or effectively reduces the  We argue that this is incompleteand compare our solutions to that case.  We also fin
% accounting for $\tau_{cor}$ involve two key factors: turbulence characterized by $\tau_{ed}$ and rotation characterized by $\Omega$.
%It was argued that rotation either leads to a decay in magnitudes and anisotropy in the $\alpha$ effect \citep{ruediger1978},
and an $\alpha\propto\Omega^{-3/4}$ for fast rotation in the fixed $\tau_{ed}$ case  (Sec \ref{2.2}). This is a weaker reduction
for fast rotators than rotational quenching in the absence of the shear effects, which predicts $\alpha\propto\Omega^{-1}$ \cite{ruediger1978}.
%EB4 i changed the above sentence to reflect your comment--

%Recently \cite{blackman2015} calls for the differential rotation as a new ingredient, since shear will shred eddies in a time scale shorter than $\tau_{ed}$ when differential rotation is strong.

%We further developed this idea by combining it with the effective $\tau_{cor}$ model caused by rotation, and making constraints on $y(Ro)=\tau_{cor}/\tau_{ed}$ by considering an upper bound of the kinetic helicity. Using (\ref{y}) and (\ref{y'}) as toy models,

Our calculations herein focused only on two specific influences  of the role of shear
and we do not purport to have captured all of the effects of shear on the turbulence  and  we have not included all terms in the EMF that depend on rotation. There are also other approaches to helicity flux driven mean field dynamos that bypass the $\alpha$ coefficient altogether. Our point in this paper however to focus  on  specific effects on shear that have been understudied.  Future work should incorporate and assess  the relevance of lessons learned here in the derivation of other dynamo coefficients not presently considered.

%Further improvements of our model exist in several aspects. The most important one is to extract a more detailed $\tau_{cor}/\tau_{ed}$, possibly from first principles. Similar procedures may also be applied to stellar systems, e.g., \cite{blackman2015}.
%caveats:
%1) no formal treatment of all terms in emf with shear
%2) assumption that MRI is same as isotropic turbulence wrt alpha
%3) not derived from first principles
%4) other approaches to helicity fluxes
%5) assume galaxy properties independent, i.e. omega and SN rate are independent

%6) observational tests, saturation and pitch angle radial dependence within given galaxy
%7) meant to be part of what would be a more genrale theory hat inclues...

%EB should we show solutions that  compare to Sur et al. to highlight differences...
\section*{Acknowledgments}{We thank  L. Chamandy and E. Vishniac for related stimulating discussions. We acknowledge support from grants HST-AR-13916.002 and NSF-AST1515648.
EB also acknowledges the Kavli Institute for Theoretical Physics (KITP) USCB
and  associated support from grant NSF PHY-1125915. }

%an $Ro$-dependent $\tau_{cor}$ using physical arguments. This is done in two separate ways, either fixing $\tau_{ed}$ or fixing $\Omega$. Adopting the 'no-z' approximation \citep{subramanian1993,moss1995,phillips2001,sur2007}, the solutions of the revised steady-state dynamo equations predict a maximum of the saturated magnetic field strength in intermediate values for $Ro$. Upon this, when we incorporate a buoyancy term caused by the Parker instability, a saturated dynamo purely supported by magnetic buoyancy becomes possible at low Rossby numbers. Other investigations also include numerical solutions of the dynamo growth time and the pitch angle.

\appendix
\section{Unified treatment of Sec. \ref{2.1} and  \ref{2.2}}\label{appx1}
We can formally combine the two cases Sec. \ref{2.1} and  \ref{2.2}
in a single formalism by defining
\beq
\tted=\frac{\tau_{ed}}{\tau_{ed0}},\ \ttr=\frac{\tau_r}{\tau_{r0}}=\frac{\Omega_0}{\Omega},
\eeq
so that $Ro=\ttr/\tted$. Then we can write
\beq
y=\tau_{cor}/\tau_{ed0}=\frac{1}{\tted^{-1}+q\ttr^{-1}}.
%EB check above
%HZ checked
\eeq
The the energy rate balance equation is
\beq
\frac{\rho v^2}{\tau_{cor}}=\frac{E}{\tau_{ed}l_{ed}^3}+\frac{0.1q^2\rho v_0^2 (\Omega/\Omega_0)^2}{\tau_s}
\eeq
%HZ 12/14
which, using $F=v/v_0$, can then be expressed as
\beq
\frac{F^2}{y}=\frac{1}{\tted^4F^3}+\frac{q^3}{10\ttr^3}.
\eeq
The solution is approximately
\beq\label{unifiedf}
F(\tted,\ttr)\simeq \text{max}\{(y/\tted^4)^{1/5},(q^3y/10\ttr^3)^{1/2}\},
\eeq
and it is related to $f$ (Eq. (\ref{f})) and $\tilf$ (Eq. (\ref{f'}))in Sec. \ref{sec2} through
\beq
f=F(\tted,1),\ \tilf=F(1,\ttr).
\eeq
%HZ 12/14 above
Non-dimensional parameters (defined in Sec. (\ref{sec3})  then exhibit the following
scalings:
\beq\label{unifiedparameter}
R_\alpha\propto y/\ttr,\ R_U\propto 1/F^{5/2}y,\ R_\omega\propto 1/F^3y\ttr,\ C\propto 1/F^3y^2.
\eeq
Fix-$\Omega$ and fix-$\tau_{ed}$ cases correspond to, respectively, taking $\ttr=1,\tted=Ro^{-1}$ and taking $\tted=1,\ttr=Ro$ in the above relations.
%EB

\bibliographystyle{mn2e}
\bibliography{RNGDNotes}
\end{document}